\documentclass[twocolumn,aps,pra,longbibliography,superscriptaddress,nofootinbib,10pt]{revtex4-1}
\usepackage[latin1]{inputenc}
\usepackage{graphicx,dcolumn,bm}
\usepackage{subfigure}
\usepackage{multirow}
\usepackage{array}
\usepackage{arydshln}
\usepackage{color}
\usepackage[colorlinks,linkcolor=blue,citecolor=blue,hyperindex]{hyperref}
\usepackage{amsfonts}
\usepackage{amssymb}
\usepackage{amsmath}
\usepackage{latexsym}
\usepackage{simplewick}
\usepackage{epstopdf}
\usepackage{multirow}
\usepackage{float}
\usepackage[table]{xcolor}
\usepackage{multibib}
\usepackage{mathrsfs}
\usepackage[sans]{dsfont}
\usepackage[mathscr]{eucal}
\usepackage{epsfig}
\usepackage{soul}
\definecolor{Blue}{rgb}{0.0,0.0,1}
\definecolor{Red}{rgb}{1,0.0,0.0}
\definecolor{Green}{rgb}{0,0.5,0.0}
\usepackage{tikz}
\usetikzlibrary{decorations.pathmorphing}
\usetikzlibrary{arrows}
\usetikzlibrary{intersections,shapes.arrows}
\usetikzlibrary{calc}
\usetikzlibrary{quotes,angles}
\usepackage{nicefrac}
\usepackage{pgfplots}
\usepgfplotslibrary{fillbetween}
\pgfplotsset{compat=1.13,colormap={violetnew}{rgb=(0.293416, 0.0574044, 0.529412) rgb=(0.394818,0.233715,0.671945) rgb =(0.49622,0.410025,0.814477) rgb=(0.588672,0.567494,0.910066) rgb=(0.663226,0.687282,0.911765) rgb=(0.73778,0.807069,0.913465) rgb=(0.807267,0.861883,0.894034) rgb=(0.874222,0.884211,0.864039) rgb=(0.941176, 0.906538, 0.834043)}}
\usepgfplotslibrary{groupplots} 
\usepgfplotslibrary[groupplots] 
\usetikzlibrary{pgfplots.groupplots} 
\usetikzlibrary[pgfplots.groupplots] 
\usepgfplotslibrary{statistics}
\usepackage{pgfplotstable}
\tikzset{jumpdot/.style={mark=*,solid},excl/.append style={jumpdot,fill=white},incl/.append style={jumpdot,fill=black}}
\begin{document}

\title{Relative purity, speed of fluctuations, and bounds on equilibration times} 
\author{Diego Paiva Pires}
\affiliation{Departamento de F\'{i}sica, Universidade Federal do Maranh\~{a}o, Campus Universit\'{a}rio do Bacanga, 65080-805, S\~{a}o Lu\'{i}s, Maranh\~{a}o, Brazil}
\author{Thiago R. de Oliveira}
\affiliation{Instituto de F\'{i}sica, Universidade Federal Fluminense, Av. Gal. Milton Tavares de Souza s/n, Gragoat\'{a}, 24210-346, Niter\'{o}i, RJ, Brazil}

\begin{abstract}
	
We discuss the local equilibration of closed systems using the relative purity, a paradigmatic information-theoretic distinguishability measure that finds applications ranging from quantum metrology to quantum speed limits. First we obtain an upper bound on the average size of the fluctuations on the relative purity: it depends on the effective dimension resembling the bound obtained with the trace distance. Second, we investigate the dynamics of relative purity and its rate of change as a probe of the speed of fluctuations around the equilibrium. In turn, such speed captures the notion of how fast some nonequilibrium state approaches the steady state under the local nonunitary dynamics, somehow giving the information of the quantum speed limit towards the equilibration. We show that the size of fluctuations depends on the quantum coherences of the initial state with respect to the eigenbasis of the Hamiltonian, also addressing the role played by the correlations between system and reservoir into the averaged speed. Finally, we have derived a family of lower bounds on the time of evolution between these states, thus obtaining an estimate for the equilibration time at the local level. These results could be of interest to the subjects of equilibration, quantum speed limits, and also quantum metrology.
\end{abstract}

\maketitle


\section{Introduction}
\label{sec:000000001}

%

Both the subjects of equilibration and thermalization fall into the very basic foundational questions of statistical mechanics: how to derive the macroscopic laws of thermodynamics from the microscopic many-particle laws. To do so, one can consider a quantum system initialized in a well defined pure state and, under some mi\-ni\-mal assumptions, still conclude the system behaves as if it were described by an equilibrium ensemble~\cite{RevModPhys.83.863}. This argument has been made technically rigorous, and also numerically confirmed by simulating several interacting quantum many-body systems~\cite{doi:10.1080/00018732.2016.1198134,PhysRevLett.110.257203}. Indeed, the problem of equilibration has attracted much interest in the last decades from both the theoretical~\cite{Eisert_Gogolin_2016} and experimental communities~\cite{PhysRevLett.89.017401,2012_NatPhys_8_267,2012_NatPhys_8_277,2012_NatPhys_8_325,Schreiber842,Kaufman794}. 

Overall, probing the mechanism of local equilibration of a closed quantum system requires answering whether and how some of its nonequilibrium states equilibrate, even if such states do not belong to a Gibbs-like statistical ensemble. The isolated system is ini\-tia\-li\-zed in a pure state and undergoes a unitary evolution governed by the time-independent Hamiltonian $H$. At the local level, the notion of equilibration involves monitoring the nonunitary dynamics of some reduced state of a small subregion of the isolated system, and quantifying how far apart it is from an equilibrium state~\cite{Short_2011}. In turn, the task of distinguishing such states can be accomplished by means of a suitable distance measure on the Hilbert space, for example the Schatten 1-norm~\cite{PhysRevE.79.061103,PhysRevE.90.012121}. So far, while there are plenty of rigorous results showing that equilibration should occur under very general conditions for small subsystems, there are a few results about the time scales involved in such physical process~\cite{TRO_2018}.

Here we will study the relative purity as a figure of merit for equilibration and show under which conditions the subsystem equilibrates. The relative purity stands as a versatile information-theoretic quantifier for distinguishing two quantum states, also being an experimental friendly measure since it relies on the overlap of density matrices~\cite{NJP_22_043001}. We also consider the rate of change of relative purity as signaling the speed of the fluctuations, thus deriving upper bounds on such velocity in a similar fa\-shion to the well-known discussion of quantum speed limits (QSLs). Noteworthy, from these inequalities we obtain lower bounds on the equilibration time of the subsystem, thus connecting both the subjects of equilibration and quantum speed limits.

The paper is organized as follows. In Sec.~\ref{sec:000000002} we review basic concepts on the subject of equilibration, also introducing the relative purity as a figure of merit to signal equilibration. In Sec.~\ref{sec:000000003} we discuss the fluctuations around the equilibrium of a bipartite closed quantum system ($\text{S}+\text{B}$), thus analyzing the dynamics of relative purity between a marginal state of subsystem $\text{S}$ and some steady state. We proved a set of lower bounds on the equilibration time that are fully characterized by the initial state and the Hamiltonian of the closed system [see Secs.~\ref{sec:000000004},~\ref{sec:000000005},~\ref{sec:000000005b},~\ref{sec:000000006}, and~\ref{sec:000000006b}]. In Sec.~\ref{sec:section000004} we illustrate our findings by means of two paradigmatic spin models, namely, the transverse field Ising model and the nonintegrable XXZ model. Finally, in Sec.~\ref{sec:conclusions} we summarize our conclusions.


\section{Relative purity and equilibration}
\label{sec:000000002}

Let us consider a quantum system described by a finite-dimensional Hilbert space $\mathcal{H} = {\mathcal{H}_{\text{S}}}\otimes{\mathcal{H}_{\text{B}}}$, with $d = \dim\mathcal{H}$, thus being split into a subsystem $\text{S}$ of dimension ${d_\text{S}} = \dim{\mathcal{H}_{\text{S}}}$, and its complement $\text{B}$ of dimension ${d_\text{B}} = \dim{\mathcal{H}_{\text{B}}}$. The whole system $\text{S}+\text{B}$ evolves unitarily under the time-independent Hamiltonian 
\begin{equation}
\label{eq:00000000001}
H = {H_{\text{S}}}\otimes{\mathbb{I}_{\text{B}}} + {\mathbb{I}_{\text{S}}}\otimes{H_{\text{B}}} + {H_{\text{SB}}} ~.
\end{equation}
The Hamiltonian is chosen to be nondegenerate and displays the spectral decomposition $H = {\sum_{n}}{E_n}|{E_n}\rangle\langle{E_n}|$, where $\{ |{E_n}\rangle \}_{n = 1,\ldots,{d_E}}$ spans an orthonormal eigenbasis related to $d_E$ distinct energy levels. In addition, we assume the energy gaps of the system are nondegenerated, i.e., ${E_i} - {E_j} \neq {E_k} - {E_l}$ for $i \neq k$ and $j \neq l$. It is worth mentioning that such assumptions may be relaxed towards degenerate systems~\cite{Reimann_2012}. The initial state of the system is pure, $\rho(0) = |{\Psi(0)}\rangle\langle{\Psi(0)}|$, and thus the instantaneous state $\rho(t) = {U}(t)\rho(0){{U}(t)^{\dagger}}$ will remain as a pure state for all times, with $U(t) = {e^{-itH}}$. For simplicity, from now on we set $\hbar = 1$. Therefore to have some kind of equilibration we have to consider a subsystem S; the rest, B, play the role of a bath. The marginal states of the quantum system are given by ${\rho_{\text{S}(\text{B})}}(t) = {\text{Tr}_{\text{B}(\text{S})}}(\rho(t))$, also written as
\begin{equation}
\label{eq:00000000002}
{\rho_{\text{S}(\text{B})}}(t) = {\sum_{j,l}}\, \langle{E_j}|\rho(0)|{E_l}\rangle \, {e^{-it({E_j} - {E_l})}} \, {\text{Tr}_{\text{B}(\text{S})}}(|{E_j}\rangle\langle{E_l}|) ~.
\end{equation}

For finite dimensional systems, it follows that $\rho_{\text{S}}(t)$ never equilibrates since it never stops to evolve; there will be recurrences. However, it may be very close to some steady state $\omega_S$ for most of the time. Let $D(x,y)$ be a suitable information-theoretic distinguishability measure of quantum states. We say subsystem equilibration has taken place at time $\tau$ when, for some $\epsilon > 0$, it follows ${\langle{D}({\rho_{\text{S}}}(t),{\omega_{\text{S}}})\rangle_T} \leq \epsilon$, for all $T > \tau$, with ${\langle{h(t)}\rangle_T} := \frac{1}{T}\,  {\int_0^T} dt \, {h}(t)$ being the time-average~\cite{PhysRevE.90.012121}. Note that, if the time average is small, then $D({\rho_{\text{S}}}(t),{\omega_{\text{S}}})$ should be small most of the time, and in this sense we say the system does equilibrate. We stress that recurrences will occur but they should be rare, and its time scale should increase with the system size.

If the equilibration process really occurs, the equilibrium state of the closed system $\text{S}+\text{B}$ is given by the infinite time-averaged state $\omega := {\langle{\rho}(t)\rangle_{\infty}} =  {\lim_{T \rightarrow \infty}} {\langle{\rho}(t)\rangle_{T}}$, and would be identical to the dephased state\footnote{Here we are not interested if the equilibrium state is a thermal or Gibbs state, which is necessary to claim that the system thermalizes. To have thermalization one need further conditions, the eigenstate thermalization hypothesis (ETH) being the most used one.}
\begin{align}
\label{eq:00000000003}
\omega&= \Delta(\rho(0)) ~, 
\end{align}
where $\Delta(\bullet) = {\sum_{j}}\, \langle{E_j}|\bullet|{E_j}\rangle|{E_j}\rangle\langle{E_j}|$ stands for the fully dephasing operator with respect to the eigenbasis of the Hamiltonian. In this setting, the steady state of subsystem $\text{S}(\text{B})$ is given by the marginal state ${\omega_{\text{S}(\text{B})}} = {\text{Tr}_{\text{B}(\text{S})}}(\omega)$. Remarkably, if one uses the Schatten 1-norm $D(x,y) := \frac{1}{2}{\|x - y\|_1}$ as a {\it bona fide} distance measure over the space of quantum states, it has been proved that 
\begin{equation}
{\lim_{T \rightarrow \infty}} {\langle{D}({\rho_{\text{S}}}(t),{\omega_{\text{S}}})\rangle_T} \leq \frac{1}{2} \sqrt{\frac{d_{\text{S}}^2}{{d^{\text{eff}}}(\omega)}} ~,
\end{equation}
where ${d^{\text{eff}}}(\omega) := 1/\text{Tr}({\omega^2})$ is the so-called effective dimension~\cite{PhysRevLett.101.190403,PhysRevE.79.061103}. More precisely, the larger the effective dimension ${d^{\text{eff}}}(\omega)$ compared to the subsystem dimension $d_{\text{S}}$, the closer the system to some steady state. Note that the effective dimension measures the number of energy eigenstates that contribute to the superposition of the initial state. It can be argued that for many-particle systems with local interactions this is typically the case, since the distance between the energy levels becomes exponentially small and it is very hard to prepare an initial state with only a few levels~\cite{PhysRevLett.101.190403}.

\begin{figure}[t]
\includegraphics[width=0.975\linewidth]{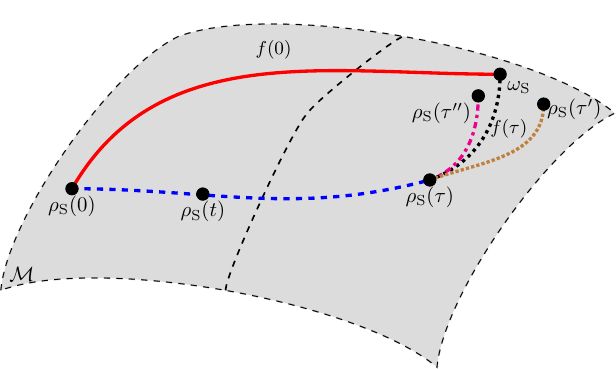}
\caption{(Color online) Depiction of the physical setting. The initial state ${\rho_{\text{S}}}(0)$ undergoes a local nonunitary evolution go\-ver\-ned by the Hamiltonian $H$ [see Eq.~\eqref{eq:00000000002}]. The evolved state ${\rho_{\text{S}}}(t)$ follows a path (blue dashed curve) over the manifold of marginal states $\mathcal{M}$ (gray surface). In practice, this state never equilibrates since it keeps evolving nonunitarily, and for instance can reach states ${\rho_{\text{S}}}(\tau)$, ${\rho_{\text{S}}}(\tau')$, or ${\rho_{\text{S}}}(\tau'')$ that are arbitrarily close to the equilibrium state $\omega_{\text{S}}$. The relative purity $f(\tau)$ (black dotted curve) captures the distinguishability between ${\rho_{\text{S}}}(\tau)$ and the equilibrium state $\omega_{\text{S}}$ of the subsystem, while $f(0)$ (red solid curve) signals how distinguishable the initial state ${\rho_{\text{S}}}(0)$ is compared to such equilibrium state. Note that, while not being a distance measure in the formal sense, the relative purity $f(t)$ provides insightful information about how far apart is the subsystem $\text{S}$ from $\omega_{\text{S}}$.}
\label{figure000000}
\end{figure}
The aforementioned criterion for equilibration is based on the closeness of states $\rho_{\text{S}}(t)$ and $\omega_{\text{S}}$ measured by the Schatten 1-norm, i.e., a geo\-me\-tric distance that signals the distinguishability between two quantum states. However, we emphasize that there are several ways to characterize such a distance~\cite{1994_PhysRevLett_72_3439}. Indeed, the convex space of quantum states is equipped with a plethora of {\it bona fide} distances~\cite{1990_Itogi_Nauki_i_Tehniki_36_69,1996_LinAlgApl_244_81,1996_JMathPhys_37_2662,Ingemar_Bengtsson_Zyczkowski}, and this nonuniqueness has been of relevance for several investigations in quantum information processing, quantum thermodynamics~\cite{1985_JournChemPhys_82_2433,PhysRevLett.125.260602}, quantum speed li\-mi\-ts~\cite{PhysRevA.67.052109,PhysRevX.6.021031}, and quantum me\-tro\-lo\-gy~\cite{IEEE_9170603_JMJM,doi:10.1116/1.5119961}, to name a few. 

Here we will consider the relative purity $f({\rho},{\varrho}) := \text{Tr}({\rho}{\varrho})$ as a natural distinguishability measure of two quantum states~\cite{PhysRevA.78.052330}. Importantly, such an information-theoretic quantifier has been useful in the study of quantum speed limits~\cite{PhysRevLett.110.050403,SciRep6_1_38149,PhysRevA.98.042132,PhysRevA.97.022109}, information scrambling and Loschmidt echoes~\cite{PhysRevLett.124.160603,PhysRevLett.124.030601}, and also for pro\-bing quantum coherence from photonic metro\-lo\-gical setups~\cite{PhysRevResearch.3.023031}. While not being a distance in the stringent sense, the re\-la\-ti\-ve purity is symmetric, $f({\rho},{\varrho}) = f({\varrho},{\rho})$, non-negative, $f({\rho},{\varrho}) \geq 0$, for all states $\rho$ and $\varrho$, and vanishes for the case in which the states are ma\-xi\-mally distin\-gui\-shable. Moreover, for ${\rho} = {\varrho}$ it recovers the quantum purity: $f({\rho},{\rho}) = \text{Tr}({\rho^2}) \leq 1$. Noteworthy, for pure states $\rho = |{\psi}\rangle\langle{\psi}|$ and $\varrho = |{\phi}\rangle\langle{\phi}|$, relative purity reduces to the pairwise fidelity, $f({\rho},{\varrho}) := {|\langle{\psi}|{\phi}\rangle|^2}$, which in turn was already investigated for understanding the equilibration of isolated thermodynamic systems~\cite{doi:10.7566_JPSJ.83.064001}. 

Overall, while one can have equilibration at the local level, the whole system never equilibrates since it evolves unitarily. In fact, in Appendix~\ref{sec:A1} we verify that the relative purity of states $\rho(t)$ and $\omega$ remains as a constant of motion of the dynamics, in turn being equal to the quantum fidelity of these states\footnote{Instead of looking at the state of the system, which can only equilibrate locally, we can look at some observables and also show they equilibrate under some general conditions. Note that even global observables, as the total magnetization, can equilibrate and in this sense one can have global equilibration.}. Hence, as for the Schatten 1-norm, we consider the relative purity $f(t) := {\text{Tr}_{\text{S}}}({\omega_{\text{S}}}\, {{\rho_{\text{S}}}(t)})$ of subsystem $\text{S}$ [see Fig.~\ref{figure000000}]. From Eqs.~\eqref{eq:00000000002} and~\eqref{eq:00000000003} it is straightforward to show that
\begin{equation}
\label{eq:00000000004}
f(t) = {\sum_{j,l}}\, \langle{E_j}|\rho(0)|{E_l}\rangle \, {e^{-it({E_j} - {E_l})}} \, \langle{E_l}|  ({\omega_{\text{S}}}\otimes{\mathbb{I}_{\text{B}}}) |{E_j}\rangle ~.
\end{equation}
For an initial nonequilibrium state, i.e., when states ${{\rho_{\text{S}}}(t)}$ and ${\omega_{\text{S}}}$ have nonoverlapping supports, one should expect the relative purity to take small values at the earlier times of the dynamics, while it approaches the infinite time-averaged value ${\langle{f(t)}\rangle_{\infty}} = {\text{Tr}_{\text{S}}}({\omega^2_{\text{S}}}) $ as the system evolves in time and equilibrates. In turn, the latter is nothing but the quantum purity of the steady state ${\omega_{\text{S}}}$ of subsystem $\text{S}$. In this way, we now introduce the following figure of merit for equilibration
\begin{equation}
\label{eq:00000000005}
g(t) := {|{f(t)} - {\langle{f(t)}\rangle_{\infty}} |^2} ~.
\end{equation}
Hence, the closer the system is to a given steady state, the smaller the figure of merit in Eq.~\eqref{eq:00000000005}, i.e., $g(t) \approx 0$ for $t > \tau$, with $\tau$ being the equilibration time. Similarly to what happens to the trace distance, it is possible to upper bound the time average of $g(t)$. In Appendix~\ref{sec:B1}, we proved that the time average of the figure of merit in Eq.~\eqref{eq:00000000005} is upper bounded as
\begin{equation}
\label{eq:00000000006}
{\langle{g(t)}\rangle_{\infty}} \leq \frac{{\|{\omega_{\text{S}}} \|_{\infty}^2}}{{d^{\text{eff}}}({\omega})} ~.
\end{equation}
Equation~\eqref{eq:00000000006} is one of the main results of the paper. It shows that the effective dimension plays a fundamental role on the equilibration process when it is monitored by the relative purity. Importantly, this result somehow agrees with the aforementioned case in which the trace distance is the distinguishability measure. The subsystem $\text{S}$ approaches the equilibrium whenever the effective dimension ${d^{\text{eff}}}({\omega})$ of the global steady state is much larger than the operator norm of the dephased marginal state $\omega_{\text{S}}$. In fact, given that ${\|{\omega_{\text{S}}}\|_{\infty}} = {\lambda_{\text{max}}}({\omega_{\text{S}}}) \leq 1$, where $\lambda_{\text{max}}(\bullet)$ sets the ma\-xi\-mum eigenvalue of the density matrix, it is reasonable to expect that ${{\|{\omega_{\text{S}}} \|_{\infty}^2}}/{{d^{\text{eff}}}({\omega})} \ll 1$ since the effective dimension of the equilibrium state $\omega$ typically takes large values. In addition, note that ${\|{\omega_{\text{S}}} \|_{\infty}^2} \leq {\|{\omega_{\text{S}}} \|_2^2}$, with ${\|{\omega_{\text{S}}} \|_2^2} = {\text{Tr}_{\text{S}}}({\omega_{\text{S}}^2}) = 1/{d^{\text{eff}}}({\omega_{\text{S}}})$, and thus the bound can be recast as ${\langle{g(t)}\rangle_{\infty}} \leq 1/[{d^{\text{eff}}}({\omega_{\text{S}}}) \, {d^{\text{eff}}}({\omega})]$. It is worth mentioning that, regardless of its simplicity, we point out the latter inequality might be less tight than the bound in Eq.~\eqref{eq:00000000006}. 


\section{Dynamics of relative purity}
\label{sec:000000003}

How fast does a given quantum system fluctuate around the equilibrium? So far, this problem has been addressed in a few works showing that such speed would be extremely small for almost all times in typical thermodynamic cases. Indeed, it can be proved that the infinity time-ave\-raged speed of state ${\rho_\text{S}}(t)$ quantified by the Schatten 1-norm reads as ${\langle {\| d{\rho_{\text{S}}}(t)/dt \|_1}\rangle_{\infty}} \leq 2\, {\| {H_{\text{S}}}\otimes{\mathbb{I}_{\text{B}}} + {H_{\text{SB}}} \|_{\infty}} \sqrt{{d^3_{\text{S}}}/ {d^{\text{eff}}}(\omega)}$~\cite{Linden_2010}, thus depending on the effective dimension and the size of the in\-te\-rac\-ting term ${H_{\text{SB}}}$ [see Eq.~\eqref{eq:00000000001}]. Furthermore, it has been shown that the speed of fluctuations around the equilibrium can be signaled by means of quantum purity, also un\-vei\-ling the role of correlations  between system and environment in the equilibration process~\cite{2010_arxiv_1003.5058}. 

Despite these remarkable theore\-ti\-cal achievements, much less is known about the time scales involved in the equilibration process~\cite{2017_arxiv_1704.06291,PhysRevX.7.031027,TRO_2018}. In this section we will investigate the dynamical behavior of re\-la\-ti\-ve purity [see Eq.~\eqref{eq:00000000004}], thus bounding its rate of change in terms of fundamental quantities such as the initial state and the Hamiltonian of the system. Furthermore, we provide bounds on the equilibration time $\tau$ in a similar fashion to the framework of quantum speed limits, i.e., the very basic question of how fast a quantum system evolves between two states.

Here we focus on the time derivative of relative purity as probing the speed of fluctuations around the equilibrium. We shall begin noticing that, since the dynamics of the subsystem $\text{S}$ is fully encoded in the diffe\-ren\-tial equation $d{\rho_{\text{S}}}(t)/dt = i {\text{Tr}_{\text{B}}}([\rho(t),H])$, with the Hamiltonian $H$ defined in Eq.~\eqref{eq:00000000001}, the absolute value of the time-derivative of relative purity becomes
\begin{equation}
\label{eq:00000000007}
\left| \frac{d}{dt}f(t) \right| = \left| i\, {\text{Tr}_{\text{SB}}} ( ({\omega_{\text{S}}}\otimes\mathbb{I}_{\text{B}}) [\rho(t),{H}] ) \right| ~.
\end{equation}
Importantly, from Eq.~\eqref{eq:00000000007} we are able to derive bounds on the rate of change of purity, which signals the fluctuations of ${\rho_{\text{S}}}(t)$ around the fixed point $\omega_{\text{S}}$ of the dynamics. From this we also derive bounds for the time in which the system approaches the equilibrium. In the following we will discuss in detail such issues.


\subsection{Relative purity and Schatten $2$-norm}
\label{sec:000000004}

Here we will show that the speed in Eq.~\eqref{eq:00000000007} satisfies an upper bound that is related to the Schatten $2$-norm and the variance of the Hamiltonian ${H}$. Let $|\text{Tr}({A_1}{A_2})| \leq {\|{A_1}\|_2}{\|{A_2}\|_2}$ be the Cauchy-Schwarz inequality for o\-pe\-ra\-tors $\{ {A_j} \}_{j = 1,2}$. In this case, Eq.~\eqref{eq:00000000007} gives rise to the following inequality
\begin{equation}
\label{eq:00000000008}
\left| \frac{d}{dt}f(t) \right| \leq {d_{\text{B}}} \,{\| {\omega_{\text{S}}} \|_2} \, {\| [\rho(t),{H}] \|_2} ~,
\end{equation}
where we have also used that ${\| {\omega_{\text{S}}}\otimes\mathbb{I}_{\text{B}} \|_2} = {d_{\text{B}}}{\| {\omega_{\text{S}}} \|_2}$. Since the time-independent Hamiltonian $H$ commutes with the evolution operator $U(t) = {e^{-itH}}$, it is straightforward to conclude that $[\rho(t),H] = {U(t)}[\rho(0),H]{U(t)^{\dagger}}$. Hence, due to the unitary invariance of the Schatten 2-norm, it follows that ${\| [\rho(t),H] \|_2} = {\| [\rho(0),H] \|_2}$, and the time average of Eq.~\eqref{eq:00000000008} over the interval $t \in [0,\tau]$ thus becomes
\begin{equation}
\label{eq:00000000010}
{\left\langle \left| \frac{d}{dt}f(t) \right|\right\rangle_{\tau}} \leq {2 \, d_{\text{B}}}  \,{\| {\omega_{\text{S}}} \|_2} \,  \sqrt{ { {\mathcal{I}_L}(\rho(0),H)} } ~,
\end{equation}
where
\begin{equation}
\label{eq:00000000011}
{\mathcal{I}_L}(\rho(0),H) := - \frac{1}{4} \, \text{Tr}({[\rho(0),H]^2}) = \frac{1}{4}\, {\| [\rho(0),H] \|_2^2} ~.
\end{equation}
We point out that the right-hand side of Eq.~\eqref{eq:00000000010} is time independent and depends on the Hamiltonian, the initial state of system $\text{S}+\text{B}$, the steady state $\omega_{\text{S}}$, and the dimension of subsystem $\text{B}$ that plays the role of a bath. Importantly, the quantifier ${\mathcal{I}_L}$ has been already introduced in the context of quantum coherence characterization, in turn defining a lower bound on the so-called Wigner-Yanase skew information~\cite{PhysRevLett.113.170401}. In this sense, the more commuting both the Hamiltonian and the initial state $\rho(0)$, the smaller the fluctuations on the speed ${\left\langle \left| {df(t)}/{dt} \right|\right\rangle_{\tau}}$.

In parti\-cu\-lar, for the pure state $\rho(0) = |\Psi(0)\rangle\langle{\Psi}(0)|$ of the global system $\text{S}+\text{B}$, the quantity ${\mathcal{I}_L}$ reduces further to half of the variance of $H$, i.e., ${\mathcal{I}_L}(\rho(0),H) = (1/2){(\Delta{H})^2} := (1/2)(\langle\Psi(0)|{H^2}|\Psi(0)\rangle - {\langle\Psi(0)|H|\Psi(0)\rangle^2})$. Next, applying the inequality $\int dx |g(x)| \geq |\int dx g(x)|$ into Eq.~\eqref{eq:00000000010}, one obtains the lower bound $\tau \geq {\tau^{(1)}}$, with
\begin{equation}
\label{eq:00000000012}
{\tau^{(1)}} := \frac{|{\text{Tr}_{\text{S}}}({\omega_{\text{S}}}\, {{\rho_{\text{S}}}(\tau)}) - {\text{Tr}_{\text{S}}}({\omega_{\text{S}}}\, {{\rho_{\text{S}}}(0)})|}{\sqrt{2} \, {d_{\text{B}}} {\| {\omega_{\text{S}}} \|_2} \, \Delta H} ~.
\end{equation}
Noteworthy, the bound in Eq.~\eqref{eq:00000000012} fits into the Mandelstam-Tamm class of quantum speed limit times for closed systems, i.e., the minimum evolution time is inversely proportional to the variance of the generator $H$~\cite{1945_JPhysURSS_9_249,Deffner2017,PhysRevX.6.021031}. If the system equilibrates at time ${\tau_{\text{eq}}}$ such as the relative purity collapses into the purity of the dephased state, i.e., ${\text{Tr}_{\text{S}}}({\omega_{\text{S}}}\, {{\rho_{\text{S}}}(\tau_{\text{eq}})}) \approx {\text{Tr}_{\text{S}}}({\omega_{\text{S}}^2})$, the lower bound in Eq.~\eqref{eq:00000000012} yields the estimation for the equilibration time as ${\tau_{\text{eq}}} \geq {\tau_{\text{eq}}^{(1)}}$, where
\begin{equation}
\label{eq:00000000013}
{\tau_{\text{eq}}^{(1)}} := \frac{\|{\omega_{\text{S}}}\|_2}{\sqrt{2} \, {d_{\text{B}}} \, \Delta H}\left|1 - \frac{{\text{Tr}_{\text{S}}}({\omega_{\text{S}}}\, {{\rho_{\text{S}}}(0)})}{{\text{Tr}_{\text{S}}}({\omega_{\text{S}}^2})}\right| ~.
\end{equation}
In particular, when ${\rho_{\text{S}}}(0)$ and $\omega_{\text{S}}$ are maximally distinguishable states, the orthogonality condition ${\text{Tr}_{\text{S}}}({\omega_{\text{S}}}\, {{\rho_{\text{S}}}(0)}) = 0$ implies the equilibration time will reduce to the case ${\tau_{\text{eq}}^{(1)}} \approx {\|{\omega_{\text{S}}}\|_2}/(\sqrt{2} \, {d_{\text{B}}} \, \Delta H)$, which will be smaller the higher the dimension of the subsystem $\text{B}$.


\subsection{Relative purity and $\ell_1$-norm of coherence}
\label{sec:000000005}

Now we will present a bound on the speed in Eq.~\eqref{eq:00000000007} that is related to the Schatten 1-norm. Let $\omega$ be the steady state of system $\text{S}+\text{B}$ that is written in terms of the eigenbasis of the Hamiltonian [see Eq.~\eqref{eq:00000000003}]. In this case, since $\omega$ and $H$ are commuting operators, i.e., $[\omega,H] = 0$, we thus have that $[\rho(t),H] = {U(t)}[\rho(0) - \omega,H]{U(t)^{\dagger}}$, where we have used the fact that ${U(t)^{\dagger}}H{U(t)} = H$, and also that ${U(t)^{\dagger}}\omega{U(t)} = \omega$ is a fixed point of the unitary dynamics. Hence, inserting this result into Eq.~\eqref{eq:00000000007} and taking its time average over the interval $t \in [0,\tau]$, one gets
\begin{equation}
\label{eq:00000000015}
{\left\langle \left| \frac{d}{dt}f(t) \right|\right\rangle_{\tau}} \leq  2 \,{\| {\omega_{\text{S}}} \|_{\infty}} {\|H\|_{\infty}} {\|\rho(0) - \omega\|_1}  ~.
\end{equation}
where we have invoked the inequality $|\text{Tr}({A_1}[{A_2},{A_3}])| \leq {\|{A_1}\|_{\infty}}{\| [{A_2},{A_3}] \|_1} \leq  2\, {\|{A_1}\|_{\infty}}{\|{A_2}\|_1}{\|{A_3}\|_{\infty}}$~\cite{BOTTCHER20081864,AUDENAERT20101126}, and employed the unitary invariance of the Schatten 1-norm, ${\| [\rho(t) - \omega,H] \|_1} = {\| [\rho(0) - \omega,H] \|_1}$, also using the identity ${\| {\omega_{\text{S}}}\otimes\mathbb{I}_{\text{B}} \|_{\infty}} = {\| {\omega_{\text{S}}} \|_{\infty}}$.

Equation~\eqref{eq:00000000015} means that the speed of fluctuations around the equilibrium is upper bounded by the pro\-duct of maximum eigenvalues of both the Hamiltonian $H$ and the steady state $\omega_{\text{S}}$. Importantly, the bound depends on the coherences of the initial state of the system. In fact, since the dephased state $\omega$ is a fully diagonal matrix that comprises the populations of $\rho(0)$ in the energy eigenbasis of $H$, the Schatten 1-norm ${\|\rho(0) - \omega\|_1}$ plays the role of the $\ell_1$-norm of coherence of $\rho(0)$ respective to such eigenbasis, thus quantifying how far apart it is from the incoherent state $\omega$~\cite{RevModPhys.89.041003}. Overall, the more incoherent the initial state with respect to the steady state, the smaller the speed of the fluctuations.

Next, applying the inequality $\int dx |g(x)| \geq |\int dx g(x)|$ into Eq.~\eqref{eq:00000000015}, one gets the lower bound $\tau \geq {\tau^{(2)}}$, with
\begin{equation}
\label{eq:00000000016}
 {\tau^{(2)}} := \frac{|{\text{Tr}_{\text{S}}}({\omega_{\text{S}}}\, {{\rho_{\text{S}}}(\tau)}) - {\text{Tr}_{\text{S}}}({\omega_{\text{S}}}\, {{\rho_{\text{S}}}(0)})|}{2\, {\| {\omega_{\text{S}}} \|_{\infty}} {\|H\|_{\infty}} {\|\rho(0) - \omega\|_1} } ~.
\end{equation}
Suppose the system equilibrates at time ${\tau_{\text{eq}}}$, with the relative purity ${\text{Tr}_{\text{S}}}({\omega_{\text{S}}}\, {{\rho_{\text{S}}}(\tau_{\text{eq}})}) \approx {\text{Tr}_{\text{S}}}({\omega_{\text{S}}^2}) = {\|{\omega_{\text{S}}}\|_2}$ recovering the purity of the steady state. In this case, given that ${\|{\omega_{\text{S}}}\|_2} \geq {\|{\omega_{\text{S}}}\|_{\infty}}$, the lower bound in Eq.~\eqref{eq:00000000016} implies that ${\tau_{\text{eq}}} \geq {\tau_{\text{eq}}^{(2)}}$, with
\begin{equation}
\label{eq:00000000017}
 {\tau_{\text{eq}}^{(2)}} := \frac{1}{2\, {\|H\|_{\infty}}  {\|\rho(0) - \omega\|_1}}  \left| 1 - \frac{{\text{Tr}_{\text{S}}}({\omega_{\text{S}}}\, {{\rho_{\text{S}}}(0)})}{{\text{Tr}_{\text{S}}}({\omega_{\text{S}}^2})}\right|  ~,
\end{equation}
We stress that, for the case of two states overlapping to zero as ${\text{Tr}_{\text{S}}}({\omega_{\text{S}}}\, {{\rho_{\text{S}}}(0)}) = 0$, Eq.~\eqref{eq:00000000017} reduces to ${\tau_{\text{eq}}^{(2)}} \approx 1/(2 \,{{\|H\|_{\infty}}  {\|\rho(0) - \omega\|_1}})$. In words, the more coherent the state $\rho(0)$ in the eigenbasis of $H$, the smaller would be $\tau_{\text{eq}}^{(2)}$, thus showing that quantum coherence of the probe state plays a role on the equilibration time.


\subsection{Relative purity and Quantum Fisher Information}
\label{sec:000000005b}

We shall point out that one may arrive at a slightly different upper bound on the speed of fluctuations by exploiting another set of inequalities. To proceed, by invo\-king both the inequality $|\text{Tr}({A_1}{A_2})| \leq {\|{A_1}\|_{\infty}}{\|{A_2}\|_1}$ and the unitary invariance ${\| [\rho(t),H] \|_1} = {\| [\rho(0),H] \|_1}$ of the Schatten 1-norm, Eq.~\eqref{eq:00000000007} is written as
\begin{equation}
\label{eq:00000000018}
\left| \frac{d}{dt}f(t) \right| \leq {\| {\omega_{\text{S}}} \|_{\infty}} {\| [\rho(0),H] \|_1} ~.
\end{equation}
Interestingly, the right-hand side of Eq.~\eqref{eq:00000000018} can be upper bounded via the inequality ${\| [\varrho,H] \|_1^2} \leq 4\, {{\mathcal{F}_Q}(\varrho,H)}$~\cite{PhysRevA.94.012339,PhysRevX.6.041044}, where ${\mathcal{F}_Q}(\varrho,H)$ is the so-called quantum Fisher information (QFI) and reads as
\begin{equation}
\label{eq:00000000019}
{\mathcal{F}_Q}(\varrho,H) = \frac{1}{2}\, {\sum_{k,l}} \frac{({\lambda_k} - {\lambda_l})^2}{{\lambda_k} + {\lambda_l}} {|\langle{k}|H|{l}\rangle|^2} ~,
\end{equation}
where $\{ {\lambda_l} , |{l}\rangle\}_l$ are the eigenvalues and eigenvectors of some mixed state $\varrho$. Hence, by inserting such bound into Eq.~\eqref{eq:00000000018}, and also taking the time average over the interval $t \in [0,\tau]$, it yields
\begin{equation}
\label{eq:00000000020}
{\left\langle \left| \frac{d}{dt}f(t) \right|\right\rangle_{\tau}} \leq  2\, {\| {\omega_{\text{S}}} \|_{\infty}} \, \sqrt{{\mathcal{F}_Q}(\rho(0),H)} ~.
\end{equation}

Equation~\eqref{eq:00000000020} means that the speed of fluctuations around the equi\-li\-brium is upper bounded by the QFI, a paradigmatic quantity in quantum metrology that is widely applied for enhance phase estimation~\cite{doi:10.1116/1.5119961,PhysRevLett.102.100401,pezze_smerzi_188_691_2014}. In particular, for the pure state $\rho(0) = |\Psi(0)\rangle\langle{\Psi(0)}|$, QFI reduces further to the variance of the generator $H$, i.e., ${\mathcal{F}_Q}(\rho(0),H) = {(\Delta{H})^2} := \langle\Psi(0)|{H^2}|\Psi(0)\rangle - {\langle\Psi(0)|H|\Psi(0)\rangle^2}$~\cite{e19030124}. It can be proved that Eq.~\eqref{eq:00000000020} implies the lower bound $\tau \geq {\tau^{(3)}}$, where
\begin{equation}
\label{eq:00000000021}
{\tau^{(3)}} := \frac{|{\text{Tr}_{\text{S}}}({\omega_{\text{S}}}\, {{\rho_{\text{S}}}(\tau)}) - {\text{Tr}_{\text{S}}}({\omega_{\text{S}}}\, {{\rho_{\text{S}}}(0)})|}{2\, {\| {\omega_{\text{S}}} \|_{\infty}}  \sqrt{{\mathcal{F}_Q}(\rho(0),H)}} ~,
\end{equation}
where we used that $\int dx |g(x)| \geq |\int dx g(x)|$. At the equilibrium, using that ${\text{Tr}_{\text{S}}}({\omega_{\text{S}}}\, {{\rho_{\text{S}}}(\tau_{\text{eq}})}) \approx {\text{Tr}_{\text{S}}}({\omega_{\text{S}}^2}) = {\|{\omega_{\text{S}}}\|_2}$, and ${\|{\omega_{\text{S}}}\|_2} \geq {\|{\omega_{\text{S}}}\|_{\infty}}$, the lower bound in Eq.~\eqref{eq:00000000021} gives rise to the following time scale for equilibration as ${\tau_{\text{eq}}} \geq {\tau_{\text{eq}}^{(3)}}$, where we define
\begin{equation}
\label{eq:00000000022}
{\tau_{\text{eq}}^{(3)}} := \frac{1}{2\Delta H}  \left|1 - \frac{{\text{Tr}_{\text{S}}}({\omega_{\text{S}}}\, {{\rho_{\text{S}}}(0)})}{{\text{Tr}_{\text{S}}}({\omega_{\text{S}}^2})}\right|  ~,
\end{equation}
which in turn will reduce to the simplest case ${\tau_{\text{eq}}^{(3)}} \approx 1/\Delta{H}$ for two maximally distinguishable states $\rho(0)$ and $\omega_{\text{S}}$.


\subsection{Relative purity and mutual information}
\label{sec:000000006}

Lastly, we obtain an upper bound that depends on the correlations between $\text{S}$ and $\text{B}$. In order to do so, we will introduce the traceless operator
\begin{equation}
\label{eq:00000000023}
\widetilde{\rho}(t) := \rho(t) - {\omega_{\text{S}}}\otimes{\omega_{\text{B}}} ~,
\end{equation}
which in turn leads to an infinity time-averaged operator ${\langle\widetilde{\rho}(t)\rangle_{\infty}} = \omega - {\omega_{\text{S}}}\otimes{\omega_{\text{B}}}$ that is a fully correlated state. The operator in Eq.~\eqref{eq:00000000023} also implies the traceless marginal states ${\widetilde{\rho}_{\text{S}}}(t) = {\text{Tr}_{\text{B}}}(\widetilde{\rho}(t)) = {\rho_{\text{S}}}(t) - {\omega_{\text{S}}}$ and ${\widetilde{\rho}_{\text{B}}}(t) = {\text{Tr}_{\text{S}}}(\widetilde{\rho}(t)) = {\rho_{\text{B}}}(t) - {\omega_{\text{B}}} $, which are also zero-valued ope\-rators under the infinity time average, i.e., ${\langle{\widetilde{\rho}_{\text{S}}}(t)\rangle_{\infty}} = {\langle{\widetilde{\rho}_{\text{B}}}(t)\rangle_{\infty}} = 0$. Next, by exploiting the cyclic property of the trace, $\text{Tr}({A_1}[{A_2},{A_3}]) = \text{Tr}([{A_1},{A_2}]{A_3})$, it is straightforward to show that ${\text{Tr}_{\text{SB}}} (({\omega_{\text{S}}}\otimes\mathbb{I}_{\text{B}}) [{\omega_{\text{S}}}\otimes{\omega_{\text{B}}},H] ) = 0$, which immediately implies the identity
\begin{equation}
\label{eq:00000000024}
{\text{Tr}_{\text{SB}}} ( ({\omega_{\text{S}}}\otimes\mathbb{I}_{\text{B}}) [\rho(t),H] ) = {\text{Tr}_{\text{SB}}} ( ({\omega_{\text{S}}}\otimes\mathbb{I}_{\text{B}}) [\widetilde{\rho}(t),H] ) ~.
\end{equation}
Inserting Eq.~\eqref{eq:00000000024} into Eq.~\eqref{eq:00000000007}, also noting that ${\text{Tr}_{\text{SB}}} (({\omega_{\text{S}}}\otimes\mathbb{I}_{\text{B}}) [\widetilde{\rho}(t),\mathbb{I}_{\text{S}}\otimes{H_{\text{B}}}] ) = 0$, one obtains
\begin{align}
\label{eq:00000000025}
\left|\frac{d}{dt}f(t) \right| &= \left| i \, {\text{Tr}_{\text{SB}}} ( ({\omega_{\text{S}}}\otimes\mathbb{I}_{\text{B}}) [\widetilde{\rho}(t),{H_{\text{S}}}\otimes{\mathbb{I}_{\text{B}}} + {H_{\text{SB}}}] ) \right|  \nonumber\\
&\leq 2\,  {\| {\omega_{\text{S}}} \|_{\infty}} {\| \widetilde{\rho}(t) \|_1}{\| {H_{\text{S}}}\otimes{\mathbb{I}_{\text{B}}} + {H_{\text{SB}}} \|_{\infty}} ~,
\end{align}
where we have used the inequalities ${|\text{Tr}({A_1}[{A_2},{A_3}])|} \leq {\|{A_1}\|_{\infty}}{\| [{A_2},{A_3}] \|_1}$, ${\| [{A_2},{A_3}] \|_1} \leq 2\, {\|{A_2}\|_1}{\|{A_3}\|_{\infty}}$~\cite{BOTTCHER20081864,AUDENAERT20101126}, and the fact that ${\| {\omega_{\text{S}}}\otimes\mathbb{I}_{\text{B}} \|_{\infty}} = {\| {\omega_{\text{S}}}\|_{\infty}}$. We point out that, from Pinsker's inequality, the trace norm of the traceless operator $\widetilde{\rho}(t)$ in Eq.~\eqref{eq:00000000023} is upper bounded as~\cite{doi:10.1063_1.4871575}
\begin{equation}
\label{eq:00000000026}
{\| {\widetilde{\rho}}(t) \|_1} \leq \sqrt{2\, S(\rho(t)\| {\omega_{\text{S}}}\otimes{\omega_{\text{B}}} )} ~,
\end{equation}
with the relative entropy defined as $S(x\|y) = - S(x) - {\text{Tr}_{\text{SB}}}(x\ln{y})$, and $S(x) = - {\text{Tr}_{\text{SB}}}(x\ln{x})$ being the von Neumann entropy. In particular, it can be proved that the relative entropy is written as $S(\rho(t)\| {\omega_{\text{S}}}\otimes{\omega_{\text{B}}}) = {I_{\text{SB}}}(\rho(t)) + S({\rho_{\text{S}}}(t)\|{\omega_{\text{S}}}) + S({\rho_{\text{B}}}(t)\|{\omega_{\text{B}}})$, and thus it depends on the correlations of the system measured by the mutual information, ${I_{\text{SB}}}(\rho(t)) := S({\rho_{\text{S}}}(t)) + S({\rho_{\text{B}}}(t)) - S(\rho(t))$. This also means the relative entropy depends on the distance $S({\rho_{\text{S},\text{B}}}(t)\|{\omega_{\text{S},\text{B}}})$ between the marginal states ${\rho_{\text{S},\text{B}}}(t)$ and $\omega_{\text{S},\text{B}}$, thus assigning a geometric perspective to the bound in Eq.~\eqref{eq:00000000026}. Inserting Eq.~\eqref{eq:00000000026} into Eq.~\eqref{eq:00000000025} and taking the time average over the interval $t \in [0,\tau]$ yields
\begin{widetext}
\begin{equation}
\label{eq:00000000028}
{{\left\langle \left| \frac{d}{dt}f(t) \right| \right\rangle}_{\tau}} \leq 2 \sqrt{2} \,  {\| {\omega_{\text{S}}} \|_{\infty}} {\| {H_{\text{S}}}\otimes{\mathbb{I}_{\text{B}}} + {H_{\text{SB}}} \|_{\infty}} \, \sqrt{ {\langle S(\rho(t)\| {\omega_{\text{S}}}\otimes{\omega_{\text{B}}} )\rangle }_{\tau} } ~,
\end{equation}
\end{widetext}
where we have exploited the concavity of the square-root function. 

Importantly, Eq.~\eqref{eq:00000000028} means that the speed of fluctuations is upper bounded by the relative entropy which distinguishes the instantaneous state of the whole system from its uncorrelated steady state, also being a function of the maximum eigenvalue of the marginal dephased state $\omega_{\text{S}}$ and the operator norm of ${H_{\text{S}}}\otimes{\mathbb{I}_{\text{B}}} + {H_{\text{SB}}}$. In this regard, it is worth noting that if the interacting Hamiltonian ${H_{\text{SB}}}$ couples the system to only a few degrees of freedom of the bath, which is typically the case of spin models with nearest-neighbor couplings, thus the upper bound in Eq.~\eqref{eq:00000000028} will be mostly independent of the size of subsystem $\text{B}$. In Appendix~\ref{sec:C1} we discussed a similar bound to the speed of fluctuations for the quantum purity for subsystem $\text{S}$.

In the limit $\tau \rightarrow \infty$, the infinity time average of the relative entropy is written as ${\langle{S}(\rho(t)\| {\omega_{\text{S}}}\otimes{\omega_{\text{B}}})\rangle_{\infty}} = S({\omega_{\text{S}}}) + S({\omega_{\text{B}}}) = 2S({\omega_{\text{S}}})$. In detail, this comes from the fact that the von Neumann entropy $S(\rho(t)) = S(\rho(0)) = 0$ remains unchanged for a pure state $\rho(t) = {U(t)}\rho(0){U^{\dagger}}(t)$ evolving unitarily, and also that the two dephased marginal states ${\omega_{\text{S},\text{B}}}$ of the bipartite system store the same amount of information, i.e., $S({\omega_{\text{S}}}) = S({\omega_{\text{B}}})$. In this case, one readily gets
\begin{equation}
\label{eq:00000000029}
{{\left\langle \left| \frac{d}{dt}f(t) \right| \right\rangle}_{\infty}} \leq 4 \,  {\| {\omega_{\text{S}}} \|_{\infty}} {\| {H_{\text{S}}}\otimes{\mathbb{I}_{\text{B}}} + {H_{\text{SB}}} \|_{\infty}}  \, \sqrt{ \ln{d_{\text{S}}} } ~,
\end{equation}
where we have used that $0 \leq S({\omega_{\text{S}}}) \leq \ln{d_{\text{S}}}$.

Next, we can make use of Eq.~\eqref{eq:00000000028} to bound the time evolution $\tau$ of the closed quantum system. Indeed, we obtain the bound $\tau \geq {\tau^{(4)}}$, where
\begin{equation}
\label{eq:00000000030}
{\tau^{(4)}} := \frac{(1/2\sqrt{2})\, |{\text{Tr}_{\text{S}}}({\omega_{\text{S}}}\, {{\rho_{\text{S}}}(\tau)}) - {\text{Tr}_{\text{S}}}({\omega_{\text{S}}}\, {{\rho_{\text{S}}}(0)})|}{ {\| {\omega_{\text{S}}} \|_{\infty}} {\| {H_{\text{S}}}\otimes{\mathbb{I}_{\text{B}}} + {H_{\text{SB}}} \|_{\infty}} \, \sqrt{ {\langle S(\rho(t)\| {\omega_{\text{S}}}\otimes{\omega_{\text{B}}}) \rangle_{\tau}} }} ~,
\end{equation}
where we have applied the inequality $|\int dx g(x)| \leq \int dx |g(x)|$. If the system approaches the equilibrium at time $\tau_{\text{eq}}$, with ${\text{Tr}_{\text{S}}}({\omega_{\text{S}}}\, {{\rho_{\text{S}}}(\tau_{\text{eq}})}) \approx {\text{Tr}_{\text{S}}}({\omega_{\text{S}}^2}) = {\|{\omega_{\text{S}}}\|_2}$, it follows that ${\tau_{\text{eq}}} \geq {\tau_{\text{eq}}^{(4)}}$, with
\begin{equation}
\label{eq:00000000031}
 {\tau_{\text{eq}}^{(4)}} := \frac{\frac{1}{2\sqrt{2}}  \left|1 - \frac{{\text{Tr}_{\text{S}}}({\omega_{\text{S}}}\, {{\rho_{\text{S}}}(0)})}{{\text{Tr}_{\text{S}}}({\omega_{\text{S}}^2})}\right|}{ {\| {H_{\text{S}}}\otimes{\mathbb{I}_{\text{B}}} + {H_{\text{SB}}} \|_{\infty}} \, \sqrt{ \text{ln}[1/{\lambda_{\text{min}}}({\omega_{\text{S}}}\otimes{\omega_{\text{B}}})] }} ~.
\end{equation}
where we used that ${\|{\omega_{\text{S}}}\|_2} \geq {\|{\omega_{\text{S}}}\|_{\infty}}$, and also invoked the inequality $S(\rho(t)\| {\omega_{\text{S}}}\otimes{\omega_{\text{B}}}) \leq \text{ln}[1/{\lambda_{\text{min}}}({\omega_{\text{S}}}\otimes{\omega_{\text{B}}})]$~\cite{doi:10.1063/1.2044667,doi:10.1063/1.3657929}, with ${\lambda_{\text{min}}}(\bullet)$ setting the minimum eigenvalue of the density matrix.


\subsection{Discussion}
\label{sec:000000006b}

In the previous sections, we presented a set of lower bounds on the speed of evolution and the equilibration time for the subsystem $\text{S}$. The relative purity signals the distinguishability between the reduced state ${\rho_{\text{S}}}(t)$ and the steady state $\omega_{\text{S}}$, thus indicating how far apart they are in the sense of witnessing whether both the states have zero or nonzero overlapping supports. From Eqs~\eqref{eq:00000000010}, and~\eqref{eq:00000000015}, we see that the average speed of the fluctuations depends on the coherences of the pure initial state $\rho(0)$ of the bipartite system. 

On the one hand, the more commuting $\rho(0)$ and $H$, the smaller the fluctuations on the speed [see Eq.~\eqref{eq:00000000010}]. However, this bound seems to be looser since it depends on the dimension $d_{\text{B}}$ of the subsystem $\text{B}$, which in turn can be large. On the other hand, the bound in Eq.~\eqref{eq:00000000015} shows that the fluctuations on the speed are constrained to the quantum fluctuations of $H$ captured by its variance regarding $\rho(0)$. Noteworthy, this bound sounds more appealing since it grows with the maximum eigenvalue of the steady state, while being of inte\-rest for metrological purposes due to its connection with the quantum Fisher information. From Eq.~\eqref{eq:00000000020}, note that the fluctuations on the speed will decrease as $\rho(0)$ approaches the equilibrium state $\omega$, which in turn is a fully incoherent state into the eigenbasis of $H$. Opposite to these results, Eq.~\eqref{eq:00000000028} depends on the correlations of the bipartite system via the existing link between relative entropy and mutual information. It is worth noting that its right-hand side is a function of time $\tau$, while the previous upper bounds are fully time independent.

The speed of the relative purity captures the notion of how fast some nonequilibrium state of subsystem $\text{S}$ approaches its steady state under the local nonunitary dynamics, somehow giving the information of the QSL towards the equilibration. In turn, such set of speeds limits implies a family of lower bounds on the time of evolution between these states. Indeed, from Eqs.~\eqref{eq:00000000012},~\eqref{eq:00000000016},~\eqref{eq:00000000021} and~\eqref{eq:00000000030}, the minimum time of evolution displays the QSL time as
\begin{equation}
\label{eq:00000000031002aaaa}
{\tau_{\text{QSL}}} := \max\{ {\tau^{(1)}}, {\tau^{(2)}}, {\tau^{(3)}}, {\tau^{(4)}} \} ~.
\end{equation}
The QSL time is a quantity that fully characterizes the dynamics of the set of eigenstates of the Hamiltonian governing the dynamics. Note that $\tau_{\text{QSL}} \equiv \tau_{\text{QSL}}(\tau)$ is a time-dependent quantity, which is expected since we are comparing states ${\rho_{\text{S}}}(\tau)$ and ${\omega_{\text{S}}}$. We point out that most of the bounds on the time evolution of different physical systems that have appeared in the literature address time-dependent QSLs (see, for example, Ref.~\cite{Deffner2017} and references therein). This ``caveat'' in the QSLs is not well discussed in the literature, and we are following the aforementioned standard procedure. From the QSL time, fixing $\rho_{\text{S}}(\tau) \approx \omega_S$, we also obtain a time scale for equilibration at the local level (this one time independent). Hence, from Eqs.~\eqref{eq:00000000013},~\eqref{eq:00000000017},~\eqref{eq:00000000022} and~\eqref{eq:00000000031}, it is possible to concatenate the previous results into a unified estimation for the equilibration time yields
\begin{equation}
\label{eq:00000000031002}
{\widetilde{\tau}_{\text{eq}}} := \max\{ {\tau^{(1)}_{\text{eq}}}, {\tau^{(2)}_{\text{eq}}}, {\tau^{(3)}_{\text{eq}}}, {\tau^{(4)}_{\text{eq}}} \} ~.
\end{equation}
We point out that the bounds in Eqs.~\eqref{eq:00000000013} and~\eqref{eq:00000000022} are inversely proportional to the variance of the Hamiltonian $H$, thus resembling the well-known class of QSLs {\it a la} Mandelstam-Tamm~\cite{Deffner2017}.


\section{Examples}
\label{sec:section000004}

\begin{figure}[t]
\includegraphics[width=1.0\linewidth]{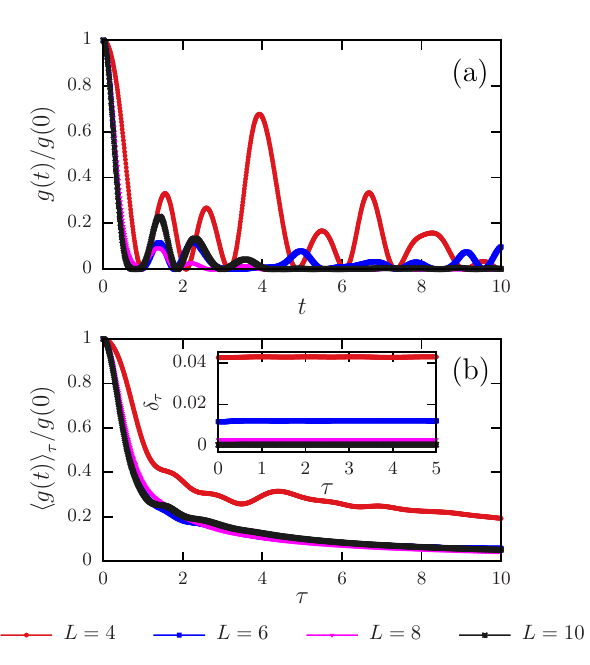}
\caption{(Color online) Plot of the figure of merit for the Ising model, with parameters $J = 1$, ${h_x} = 0.5$, ${h_z} = -1.05$ [see Eq.~\eqref{eq:00000000032}]. Here, we set the system sizes $L = \{4,6,8,10\}$, with open boun\-da\-ry conditions, while $L_S = 1$, and $L_B = \{3,5,7,9\}$. The system is initialized at the charge-density-wave-like state $|{\Psi(0)}\rangle = |1,0,1,0,\ldots,0,1\rangle$, with $|0\rangle$ and $|1\rangle$ denoting the spin-up and -down state, respectively.}
\label{figure000001}
\end{figure}
%
\begin{figure}[t]
\includegraphics[width=1.0\linewidth]{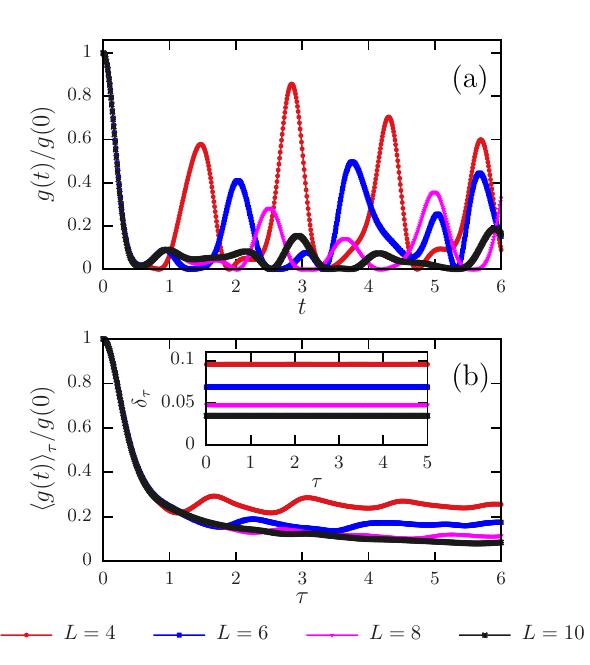}
\caption{(Color online) Plot of the normalized figure of merit for the non-integrable XXZ model, with parameters $J = 1$, $U = 2$, ${J_{nnn}} = 0.2$ [see Eq.~\eqref{eq:00000000033}]. Here, we set the system sizes $L = \{4,6,8,10\}$, with open boun\-da\-ry conditions, while $L_S = 1$, and $L_B = \{3,5,7,9\}$. The system is initialized at the charge-density-wave-like state $|{\Psi(0)}\rangle = |1,0,1,0,\ldots,0,1\rangle$, with $|0\rangle$ and $|1\rangle$ denoting the spin-up and -down state, respectively.}
\label{figure000002}
\end{figure}

In the following we will illustrate our findings by focusing on two prototypical quantum many-body systems. The first is the transverse field Ising model with local fields:
\begin{equation}
\label{eq:00000000032}
{H_{\text{Ising}}} = J{\sum_{j = 1}^{L - 1}}\, {\sigma_j^x}{\sigma_{j + 1}^x} + {\sum_{j = 1}^L}\left({h_x}{\sigma_j^x} + {h_z}{\sigma_j^z}\right) ~,
\end{equation}
with parameters $J = 1$, ${h_x} = 0.5$, ${h_z} = -1.05$~\cite{2017_arxiv_1704.06291}. The equilibration properties of this model have already been numerically investigated, particularly identifying initial states and sets of parameters for which equilibration occurs rapidly~\cite{leviatan2017quantum}, or even never takes place~\cite{PhysRevLett.106.050405}. The se\-cond is the nonintegrable XXZ model with next-nearest-neighbor hopping:
\begin{align}
\label{eq:00000000033}
{H_{\text{XXZ}}} &= J{\sum_{j = 1}^{L - 1}}\left( {\sigma_j^x}{\sigma_{j + 1}^x} + {\sigma_j^y}{\sigma_{j + 1}^y}\right)  + U {\sum_{j = 1}^L}\, {\sigma_j^z}{\sigma_{j + 1}^z}  \nonumber\\
&+  {J_{nnn}}{\sum_{j = 1}^{L - 2}}\left( {\sigma_j^x}{\sigma_{j + 1}^z}{\sigma_{j + 2}^x} + {\sigma_j^y}{\sigma_{j + 1}^z}{\sigma_{j + 2}^y}\right) ~,
\end{align}
where we set the input configuration $J = 1$, $U = 2$, and ${J_{nnn}} = 0.2$~\cite{2017_arxiv_1704.06291}. The two spin models are initialized in a charge-density-wave-like state, i.e., $|{\Psi(0)}\rangle = |1,0,1,0,\ldots,0,1\rangle$, with $|0\rangle$ and $|1\rangle$ denoting the spin-up and -down state, respectively. Here we will investigate the role played by the figure of merit $g(t)$ [see Eq.~\eqref{eq:00000000005}] for signaling the equilibration process in both many-body quantum systems. We will also discuss the tightness of the bound in Eq.~\eqref{eq:00000000006} by introducing the relative error
\begin{equation}
\label{eq:00000000034}
{\delta_{\tau}} := \frac{{\|{\omega_{\text{S}}} \|_{\infty}^2}}{{d^{\text{eff}}}({\omega})} - {\langle{g(t)}\rangle_{\tau}} ~.
\end{equation}
Overall, the smaller the relative error, the tighter the bound on the fluctuations of the relative purity captured by the function $g(t)$. In general, it is reasonable to expect that ${\lim_{\tau\rightarrow\infty}} \, {\delta_{\tau}} \approx 0$ as we increase the system size $L$ of the system. From now on we set the system sizes $L = \{4,6,8,10\}$, where $L_S = 1$, and $L_B = \{3,5,7,9\}$.

In Fig.~\ref{figure000001}, we show plots of the figure of merit $g(t)$ [see Eq.~\eqref{eq:00000000005}] for the Ising model with open boun\-da\-ry conditions. In Fig.~\ref{figure000001}(a), we plot the normalized time signal $g(t)/g(0)$ as a function of time. Noteworthy, the recurrences exhibited in the signal are mostly suppressed as we increase the system size $L$. In other words, we expect that the fluctuations tend to decrease in the limit of larger system sizes. In Fig.~\ref{figure000001}(b), we plot the finite time average of the figure of merit, ${\langle {g(t)} \rangle_{\tau}}/g(0)$. Next, in Fig.~\ref{figure000002} we show our results for the XXZ model with open boun\-da\-ry conditions. In Fig.~\ref{figure000002}(a), we show the plots of the normalized time signal $g(t)/g(0)$, while Fig.~\ref{figure000002}(b) show the plot of ${\langle {g(t)} \rangle_{\tau}}/g(0)$. The results are quite similar to the case of the Ising model. Overall, note the size of fluctuations in $g(t)$ decreases as we increase the system size $L$, thus signaling the system equilibrates.

The insets in Figs.~\ref{figure000001}(b) and~\ref{figure000002}(b) show the plot of the relative error ${\delta_{\tau}}$ [see Eq.~\eqref{eq:00000000034}] as a function of time $\tau$. In agreement with Eq.~\eqref{eq:00000000006}, the relative error satisfies the condition $\delta_{\tau} \geq 0$ for all $\tau \geq 0$. In addition, the amplitude of the relative error decreases as the system size increases. We see that each of the plots saturates at fixed values for all times. To see this in detail, we first note that ${\langle{g(t)}\rangle_{\tau}}$ is a time-dependent function, while the quantity ${{\|{\omega_{\text{S}}} \|_{\infty}^2}}/{{d^{\text{eff}}}({\omega})}$ is time independent and stands as a constant for a given system size $L$. We point out that the time-averaged quantity ${\langle{g(t)}\rangle_{\tau}} $ is smaller than the ratio ${{\|{\omega_{\text{S}}} \|_{\infty}^2}}/{{d^{\text{eff}}}({\omega})}$ by some orders of magnitude, for all $\tau \geq 0$ and system size $L$. In other words, the time oscillations of ${\langle{g(t)}\rangle_{\tau}}$ are negligible when compared to the constant value of ${{\|{\omega_{\text{S}}} \|_{\infty}^2}}/{{d^{\text{eff}}}({\omega})}$.

\begin{figure}[t]
\includegraphics[width=1.0\linewidth]{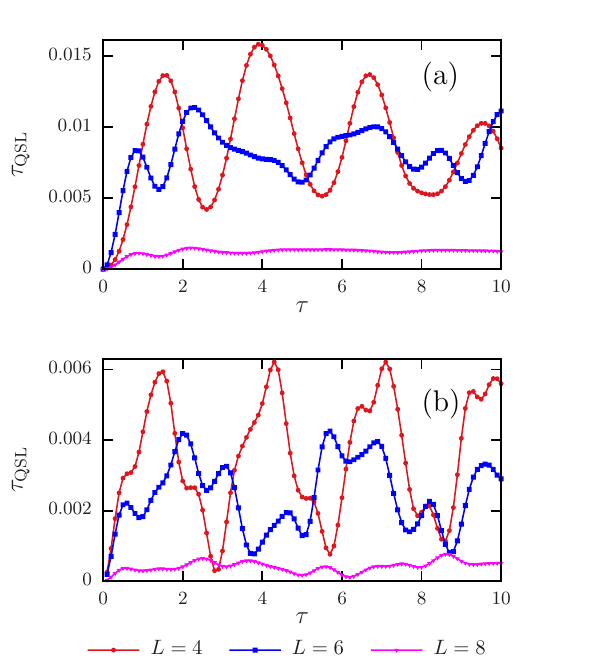}
\caption{(Color online) Plot of the QSL time ${\tau}_{\text{QSL}}$ [see Eq.~\eqref{eq:00000000031002aaaa}] for (a) the Ising model, with parameters $J = 1$, ${h_x} = 0.5$, ${h_z} = -1.05$ [see Eq.~\eqref{eq:00000000032}], and (b) the nonintegrable XXZ model, with parameters $J = 1$, $U = 2$, ${J_{nnn}} = 0.2$ [see Eq.~\eqref{eq:00000000033}]. Here, we set the system sizes $L = \{4,6,8\}$, with open boun\-da\-ry conditions, while $L_S = 1$, and $L_B = \{3,5,7\}$. The system is initialized at the charge-density-wave-like state $|{\Psi(0)}\rangle = |1,0,1,0,\ldots,0,1\rangle$, with $|0\rangle$ and $|1\rangle$ denoting the spin-up and -down state, respectively.}
\label{figure000003}
\end{figure}

In Fig.~\ref{figure000003}, we plot the QSL time in Eq.~\eqref{eq:00000000031002aaaa} for both the Ising model [see Fig.~\ref{figure000003}(a)] and XXZ model [see Fig.~\ref{figure000003}(b)]. Note that the QSL time exhibits nonperiodic oscillations the amplitudes of which are suppressed as we increase the system size. From Figs.~\ref{figure000003}(a) and~\ref{figure000003}(b), we see that the larger the system size $L$, the smaller the amplitude of the QSL time. We find that, regardless of the time-dependent be\-ha\-vior of $\tau^{(1)}$, $\tau^{(2)}$, and $\tau^{(4)}$, it follows that $\tau^{(3)}$ dominates the numerical maximization indicated in Eq.~\eqref{eq:00000000031002aaaa}. We refer to Appendix~\ref{sec:D1} for more details on the set of QSL times. Note that $\tau^{(3)}$ is inversely proportional to the variance $\Delta{H}$ [see Eq.~\eqref{eq:00000000021}]. Given the initial pure state $|{\Psi(0)}\rangle = |1,0,1,0,\ldots,0,1\rangle$, the variance of the Ising Hamiltonian in Eq.~\eqref{eq:00000000032} read as $\Delta{H_{\text{Ising}}} = \sqrt{ (L - 1){J^2} + L{h_x^2}}$, while one gets the variance $\Delta{H_{\text{XXZ}}} = 2J\sqrt{L - 1}$ for the XXZ model in Eq.~\eqref{eq:00000000033} with respect to such initial state. For larger system size $L$, these variances become $\Delta {H} \sim {L^{1/2}}$, and in this case the QSL time will behave as ${\tau}_{\text{QSL}} = {\tau^{(3)}} \sim (1/2) {L^{-1/2}} \, {\| {\omega_{\text{S}}} \|_{\infty}^{-1}} \, {|{\text{Tr}_{\text{S}}}({\omega_{\text{S}}}\, {{\rho_{\text{S}}}(\tau)}) - {\text{Tr}_{\text{S}}}({\omega_{\text{S}}}\, {{\rho_{\text{S}}}(0)})|}$. We expect this result should hold in the limit of larger va\-lues of $L$, but it can already be seen in Fig.~\ref{figure000003} that the amplitude of the QSL time decreases as the system size $L$ grows.

Next, we comment on the equilibration time $\widetilde{\tau}_{\text{eq}}$ in Eq.~\eqref{eq:00000000031002}. For both the aforementioned spin systems, we find that $\widetilde{\tau}_{\text{eq}} = \tau^{(3)}_{\text{eq}} = (1/2){(\Delta H)^{-1}} |1 - {{\text{Tr}_{\text{S}}}({\omega_{\text{S}}}\, {{\rho_{\text{S}}}(0)})}{[ {\text{Tr}_{\text{S}}}({\omega_{\text{S}}^2})]^{-1}} |$. For the transverse field Ising model we find the equilibration time ${\widetilde{\tau}_{\text{eq}}} \approx 9\times{10^{-3}}$ ($L = 4$ and $6$), and ${\widetilde{\tau}_{\text{eq}}} \approx {10^{-3}}$ ($L = 8$). For the nonintegrable XXZ model, the equilibration time reads as ${\widetilde{\tau}_{\text{eq}}} \approx 3\times{10^{-3}}$ ($L = 4$ and $6$), and ${\widetilde{\tau}_{\text{eq}}} \approx 4\times{10^{-4}}$ ($L = 8$). In both cases, ${\widetilde{\tau}_{\text{eq}}}$ has units of $\hbar/J$. In the limit of large system size $L$, since the variances behave as $\Delta H \sim {L^{1/2}}$, thus Eqs.~\eqref{eq:00000000022} and~\eqref{eq:00000000031002} imply that $\widetilde{\tau}_{\text{eq}} = \tau^{(3)}_{\text{eq}} \sim (1/2){L^{-1/2}} \, \left|1 - {{\text{Tr}_{\text{S}}}({\omega_{\text{S}}}\, {{\rho_{\text{S}}}(0)})}{[ {\text{Tr}_{\text{S}}}({\omega_{\text{S}}^2})]^{-1}} \right|$. Hence, again we expect the lower bound on the equilibration time to scale with $1/\sqrt{L}$ for large $L$, but already see evidence of such decay for the small values of $L$ we used.

We close this section discussing the equilibration time and the lower bound obtained on both spin systems. In Figs.~\ref{figure000001} and~\ref{figure000002}, we see that the system starts to equilibrate at a time $\tau$ such that the figure of merit $g(\tau)$ approaches a zero value. On the one hand, for the transverse field Ising model, Fig.~\ref{figure000001}(a) shows that this time is of order $\tau \approx 1$ for most of the system sizes, while one gets $\widetilde{\tau}_{\text{eq}} \approx 10^{-3}$. On the other hand, for the nonintegrable XXZ model, Fig.~\ref{figure000001}(b) shows that $\tau \approx 0.5$, but we have found the earlier times ${10^{-4}} \lesssim {\widetilde{\tau}_{\text{eq}}} \lesssim {10^{-3}}$ for the referred system sizes. In both cases, we see that the bound is fulfilled ($\tau \geq {\widetilde{\tau}_{\text{eq}}}$), but is not tight for these two spin models and the initial state we choose. While the bound may be tight for other models for larger $L$, we offer some reasons for it not being tight.

The QSL bound is obtained bounding from above the time derivative of the relative purity $f(t)$; we are trying to obtain the minimum time for a change in $f(t)$ assuming it always varies at its maximum rate [see Sec.~\ref{sec:000000003}]. In this sense, the bound is expected to not be tight if $f(t)$ strongly oscillates. In fact, the tightness of the QSL is related to the distinguishability measure between quantum states. Tighter QSL bounds have been discussed for information-theoretic quantifiers such as the quantum Fisher information~\cite{PhysRevLett.110.050402,arxiv_2108.04261}, Wigner-Yanase skew information~\cite{PhysRevX.6.021031}, and also geometric measures based on the Bures angle~\cite{PhysRevLett.120.060409}. Furthermore, we have invoked several inequalities to derive those lower bounds. In spite of simplifying the calculations, applying such inequalities may have compromised the tightness of the bounds.
	
Finally, although the bound does not provide quantitative information about the equilibration time for the spin models and initial state used as examples, they still provide insight into the physical properties involved in the equilibration process [see Sec.~\ref{sec:000000006b}]. They are also of interest, since they allow one to connect both the subjects of equilibration and speed limits. Lastly, we also showed that the relative purity is a useful witness for equilibration, and it has the advantage of being more amenable to analytical calculation and experimental measure than other figures of merit as the trace distance, for example. We emphasize that the lower bounds on the equilibration time could be tightened by invoking some minimal amount of inequalities, and also verifying other distinguishability measures. Indeed, this is an issue that we hope to address in further investigations.

\section{Conclusions}
\label{sec:conclusions}

In conclusion, we have discussed the local equilibration of closed quantum systems and the speed of fluctuations around the equilibrium. We provided a criterion for witnessing equilibration at the local level by introducing a figure of merit that is rooted in relative purity [see Eq.~\eqref{eq:00000000006}]. In turn, the latter stands as a distinguishability measure of quantum states, particularly quantifying the overlap between a nonequilibrium state of a small subsystem and a given steady state. We show that the relative purity is a useful witness for equilibration, and it has the advantage of being more amenable to analytical calculation and experimental measure than other figures of merit as the trace distance, for example. We have proved an upper bound on such figure of merit that depends on the effective dimension of the equilibrium state of the closed system. We find that the larger the effective dimension, the smaller the size of fluctuations around the system. Indeed, this somehow agrees with previous results reported in the literature where the authors have considered the Schatten 1-norm as a {\it bona fide} measure for equilibration.

We have analyzed the dynamics of relative purity and its rate of change as a probe of the speed of fluctuations around equilibrium. Indeed, we have proved a set of upper bounds on such averaged speed that depends on the initial state and the Hamiltonian of the isolated system [see Eqs.~\eqref{eq:00000000010},~\eqref{eq:00000000015},~\eqref{eq:00000000020}, and~\eqref{eq:00000000029}]. Overall, the bounds show that such fluctuations depend on the co\-herences of the pure initial state regarding the re\-fe\-ren\-ce eigenbasis of the Hamiltonian. Furthermore, we show the averaged speed also depends on the correlations of the bipartite system that are quantified via the relative entropy and mutual information. We find that if the interacting Hamiltonian ${H_{\text{SB}}}$ couples the system to only a few degrees of freedom of the bath, which is the case of spin models with nearest-neighbor couplings, such upper bound does not scale with the size of the subsystem $\text{B}$.

From these speeds we have derived a family of lower bounds on the time of evolution between such states, thus obtaining an estimate for the equilibration time at the local level. Importantly, regarding the equilibration time, we follow the same procedure as the one that is commonly applied in the derivation of quantum speed limits [see Eqs.~\eqref{eq:00000000013},~\eqref{eq:00000000017},~\eqref{eq:00000000022},~\eqref{eq:00000000031}, and~\eqref{eq:00000000031002}]. We have verified the bounds on the equilibration time are not tight for the spin models and initial state used as examples, but they still provide insight in the physical properties involved in the equilibration process. Indeed, some of the bounds fit into the Mandelstam-Tamm class of QSLs due to its dependence on the inverse of the variance of the Hamiltonian. Hence, our results somehow may bridge both the subjects of QSLs and equilibration. Finally, we believe that our results may find applications in the study of equilibration of many-body quantum systems and quantum speed limits, also being useful for discussing the enhancing of phase estimation in quantum systems around equilibrium that is of interest to quantum metro\-lo\-gy.


\begin{acknowledgments}
This work is supported by the Brazilian National Institute of Science and Technology for Quantum Information (INCT-IQ) Grant No. 465469/2014-0, and by the Air Force Office of Scientific Research under Grant No. FA9550-19-1-0361.
\end{acknowledgments}

\setcounter{equation}{0}
\setcounter{table}{0}
\setcounter{section}{0}
\numberwithin{equation}{section}
\makeatletter
\renewcommand{\thesection}{\Alph{section}} 
\renewcommand{\thesubsection}{\Alph{section}.\arabic{subsection}}
\def\@gobbleappendixname#1\csname thesubsection\endcsname{\Alph{section}.\arabic{subsection}}
\renewcommand{\theequation}{\Alph{section}\arabic{equation}}
\renewcommand{\thefigure}{\arabic{figure}}
\renewcommand{\bibnumfmt}[1]{[#1]}
\renewcommand{\citenumfont}[1]{#1}

\section*{Appendix}


\section{Relative purity and Uhlmann fidelity}
\label{sec:A1}

In this Appendix we will show that both the relative purity and Uhlmann fidelity stand as constants of motions with respect to the global unitary evolution, also taking identical values for states $\rho(t)$ and $\omega_{\text{S}}$. Let $\rho(0) = |\psi(0)\rangle\langle\psi(0)|$ be the initial state of the system $\text{S}+\text{B}$ that undergoes the unitary evolution $\rho(t) = {U}(t)\rho(0){U^{\dagger}}(t)$, with $U(t) = {e^{-itH}}$ being the evolution operator, and $H = {\sum_j}\, {E_j}|{E_j}\rangle\langle{E_j}|$ being the time-independent Hamiltonian of the system. In turn, ${\omega} = \langle{\rho(t)}\rangle_{\infty}$ is the infinite time-averaged state of the full system, also written in the energy eigenbasis of $H$ as
\begin{equation}
\label{eq:A00000000001}
{\omega} = {\sum_j}\, \langle{E_j}|\rho(0)|{E_j}\rangle|{E_j}\rangle\langle{E_j}| ~.
\end{equation}
From Eq.~\eqref{eq:A00000000001}, note that both the Hamiltonian $H$ and the dephased state $\omega$ are commuting operators, thus implying that ${U^{\dagger}}(t)\omega{U}(t) = \omega$. In this case, it is straightforward to conclude the relative purity $F(\rho(t),{\omega}) = \text{Tr}(\omega\rho(t))$ of such states is given by
\begin{align}
\label{eq:A00000000002}
F(\rho(t),{\omega}) &= \text{Tr}(\omega{U}(t)\rho(0){U^{\dagger}}(t)) \nonumber\\
&= \text{Tr}({U^{\dagger}}(t)\omega{U}(t)\rho(0)) \nonumber\\
&= \text{Tr}(\omega\rho(0)) ~.
\end{align}
Clearly, Eq.~\eqref{eq:A00000000002} shows that $F(\rho(t),{\omega}) = F(\rho(0),{\omega}) = \text{Tr}(\omega\rho(0))$ stands as a time-independent quantity that depends on the initial state of the full system and also its steady state.

Next, the Uhlmann fidelity of states $\rho(t)$ and $\omega$ is written as~\cite{Ingemar_Bengtsson_Zyczkowski} 
\begin{equation}
\label{eq:A00000000003}
\widetilde{F}(\rho(t),{\omega}) = {\left(\text{Tr}\left[\sqrt{\sqrt{\rho(t)} \, \omega\sqrt{\rho(t)}} \,\, \right]\right)^2} ~.
\end{equation}
We stress that Uhlmann fidelity is a positive quantity for all quantum states, also being a symmetric function over its entries, i.e., $\widetilde{F}(\rho(t),{\omega}) = \widetilde{F}({\omega},\rho(t))$. Invoking the identity $\sqrt{\rho(t)} = {U(t)}\sqrt{\rho(0)}{U^{\dagger}}(t)$, which holds for any density matrix undergoing a given unitary evolution~\cite{Bathia_Rajendra,PhysRevA.91.042330}, and since ${U^{\dagger}}(t)\,\omega {U}(t) = \omega$, one gets
\begin{align}
\label{eq:A00000000004}
\sqrt{\rho(t)} \, \omega\sqrt{\rho(t)} &= {U(t)}\sqrt{\rho(0)} \, \omega\sqrt{\rho(0)}\, {U^{\dagger}}(t) \nonumber\\
&=  {\langle\psi(0)|\omega|\psi(0)\rangle}\,\rho(t) ~,
\end{align}
where we have also used the fact that $\sqrt{\rho(0)} = \rho(0) = |\psi(0)\rangle\langle\psi(0)|$ since the initial state is pure. Hence, by plugging Eq.~\eqref{eq:A00000000004} into Eq.~\eqref{eq:A00000000003}, one readily gets
\begin{equation}
\label{eq:A00000000005}
\widetilde{F}(\rho(t),{\omega}) = \langle\psi(0)|\omega|\psi(0)\rangle ~.
\end{equation}
Analogously to the case of relative purity, this means the Uhlmann fidelity stands as a time-independent quantity, also being a function of both the initial and equilibrated states. Indeed, from Eq.~\eqref{eq:A00000000002}, we point out that $\widetilde{F}(\rho(t),{\omega})$ is no\-thing but the relative purity $F(\rho(0),\omega)$, and it yields
\begin{align}
\label{eq:A00000000006}
\widetilde{F}(\rho(t),{\omega}) &= \text{Tr}(\omega\, |\psi(0)\rangle \langle\psi(0)|) \nonumber\\
&= \text{Tr}(\omega\rho(0)) \nonumber\\
&= F(\rho(t),{\omega}) ~.
\end{align}
As a final remark, we emphasize the identities presented in Eqs.~\eqref{eq:A00000000002} and~\eqref{eq:A00000000005} come from the fact that $\omega$ is a fixed point of the unitary dynamics of the system, thus implying that the relative purity and Uhlmann fidelity characterize such a constant of motion.


\section{Bound on the figure of merit}
\label{sec:B1}

In this Appendix we will present in detail the derivation of the inequality presented in Eq.~\eqref{eq:00000000006}, which in turn stands as an upper bound on the figure of merit for equilibration given by
\begin{equation}
\label{eq:B00000000001}
g(t) = {\left| {\text{Tr}_{\text{S}}}\left[({\rho_{\text{S}}}(t) - {\omega_{\text{S}}}) \, {\omega_{\text{S}}}\right] \right|^2} ~.
\end{equation}
From Eqs.~\eqref{eq:00000000003} and~\eqref{eq:00000000005}, it is straightforward to conclude that
\begin{equation}
\label{eq:B00000000002}
{\rho_{\text{S}}}(t) - {\omega_{\text{S}}} = {\sum_{k \neq l}} \, {c_k}{c_l^*} \, {e^{-it({E_k} - {E_l})}} \, {\text{Tr}_{\text{B}}}(|{E_k}\rangle\langle{E_l}|) ~,
\end{equation}
where we have defined ${c_j} := \langle{E_j}|\psi(0)\rangle$. From Eq.~\eqref{eq:B00000000002}, the time average of the figure of merit in Eq.~\eqref{eq:B00000000001} becomes
\begin{widetext}
\begin{equation}
\label{eq:B00000000003}
{\langle{g(t)}\rangle_{\infty}} = {\sum_{k \neq l}} \, {\sum_{m \neq n}} \, {c_k}{c_l^*}{c_m}{c_n^*} \, {\left\langle{e^{-it({E_k} - {E_l} + {E_m} - {E_n})}}\right\rangle_{\infty}} \, {\text{Tr}_{\text{SB}}}\left[|{E_k}\rangle\langle{E_l}|({\omega_{\text{S}}}\otimes{\mathbb{I}_{\text{B}}})\right] {\text{Tr}_{\text{SB}}}\left[|{E_m}\rangle\langle{E_n}|({\omega_{\text{S}}}\otimes{\mathbb{I}_{\text{B}}})\right] ~.
\end{equation}
To evaluate the time average in the right-hand side of Eq.~\eqref{eq:B00000000003}, we will use the fact that the Hamiltonian $H$ has nondegenerate energy gaps, and thus the double summation will only include terms where $k\neq l$ and $m \neq n$~\cite{PhysRevE.79.061103}. In this case, we find the only nonzero terms are those matrix elements labeled as $n = k$ and $m = l$, thus implying that
\begin{align}
\label{eq:B00000000004}
{\langle{g(t)}\rangle_{\infty}} &= {\sum_{k \neq l}} \, {|{c_k}|^2}{|{c_l}|^2} \, {\text{Tr}_{\text{SB}}}\left[|{E_k}\rangle\langle{E_l}|({\omega_{\text{S}}}\otimes{\mathbb{I}_{\text{B}}})\right] {\text{Tr}_{\text{SB}}}\left[|{E_l}\rangle\langle{E_k}|({\omega_{\text{S}}}\otimes{\mathbb{I}_{\text{B}}})\right] \nonumber\\
&=  {\sum_{k \neq l}} \, {|{c_k}|^2}{|{c_l}|^2} \, \langle{E_l}|\,{\omega_{\text{S}}}\otimes{\mathbb{I}_{\text{B}}}|{E_k}\rangle \langle{E_k}|\, {\omega_{\text{S}}}\otimes{\mathbb{I}_{\text{B}}}|{E_l}\rangle ~,
\end{align}
where from the first to the second line we have used the cyclic property of trace. We point out that the sum in Eq.~\eqref{eq:B00000000004} can be recast as ${\sum_{k \neq l}} = {\sum_{k, l}} - {\sum_{k = l}}$, and thus one gets
\begin{align}
\label{eq:B00000000005}
{\langle{g(t)}\rangle_{\infty}} &=  {\sum_{k, l}} \, {|{c_k}|^2}{|{c_l}|^2} \, \langle{E_l}|\, {\omega_{\text{S}}}\otimes{\mathbb{I}_{\text{B}}}|{E_k}\rangle \langle{E_k}|\, {\omega_{\text{S}}}\otimes{\mathbb{I}_{\text{B}}}|{E_l}\rangle - {\sum_{k}} \, {|{c_k}|^4}\, {\langle{E_k}|\, {\omega_{\text{S}}}\otimes{\mathbb{I}_{\text{B}}}|{E_k}\rangle^2} \nonumber\\
&=  {\text{Tr}_{\text{SB}}}\left[ {\omega} ({\omega_{\text{S}}}\otimes{\mathbb{I}_{\text{B}}}) {\omega} ({\omega_{\text{S}}}\otimes{\mathbb{I}_{\text{B}}}) \right]  - {\sum_{k}} \, {|{c_k}|^4}\, {\langle{E_k}|\, {\omega_{\text{S}}}\otimes{\mathbb{I}_{\text{B}}}|{E_k}\rangle^2} \nonumber\\
&\leq {\text{Tr}_{\text{SB}}}\left[ {\omega} ({\omega_{\text{S}}}\otimes{\mathbb{I}_{\text{B}}}) {\omega} ({\omega_{\text{S}}}\otimes{\mathbb{I}_{\text{B}}}) \right] ~,
\end{align}
\end{widetext}
where we have recognized the equilibrium state $\omega = {\sum_j}{|{c_j}|^2}|{E_j}\rangle\langle{E_j}|$ [see Eq.~\eqref{eq:00000000003}], and also used that ${\sum_{k}} \, {|{c_k}|^4}\, {\langle{E_k}|\, {\omega_{\text{S}}}\otimes{\mathbb{I}_{\text{B}}}|{E_k}\rangle^2} \geq 0$. Invoking the Cauchy-Scharwz inequality for operators, i.e., $|\text{Tr}(AB)| \leq {\|{A}\|_2} {\|{B}\|_2}$, and choosing $A = B = {\omega} ({\omega_{\text{S}}}\otimes{\mathbb{I}_{\text{B}}})$, one may ve\-ri\-fy Eq.~\eqref{eq:B00000000005} implies that
\begin{equation}
\label{eq:B00000000006}
{\langle{g(t)}\rangle_{\infty}} \leq {\text{Tr}_{\text{SB}}}\left[ ({\omega_{\text{S}}}\otimes{\mathbb{I}_{\text{B}}})\,{\omega^2} ({\omega_{\text{S}}}\otimes{\mathbb{I}_{\text{B}}}) \right] ~.
\end{equation}
Next, given that $\text{Tr}(PQ) \leq {\|P\|_{\infty}}\text{Tr}(Q)$ for two positive operators $P$ and $Q$, it follows that
\begin{align}
\label{eq:B00000000007}
{\langle{g(t)}\rangle_{\infty}} &\leq {\|{\omega_{\text{S}}}\otimes{\mathbb{I}_{\text{B}}} \|_{\infty}} {\text{Tr}_{\text{SB}}}\left[ {\omega^2} ({\omega_{\text{S}}}\otimes{\mathbb{I}_{\text{B}}}) \right] \nonumber\\
&\leq {\|{\omega_{\text{S}}}\otimes{\mathbb{I}_{\text{B}}} \|_{\infty}^2} {\text{Tr}_{\text{SB}}}\left( {\omega^2} \right) ~.
\end{align}
Note that Eq.~\eqref{eq:B00000000007} can be recast by using that ${\|{\omega_{\text{S}}}\otimes{\mathbb{I}_{\text{B}}} \|_{\infty}} = {\|{\omega_{\text{S}}} \|_{\infty}}$, and also recognizing the effective dimension ${d^{\text{eff}}}({\omega}) = 1/{\text{Tr}_{\text{SB}}}\left( {\omega^2} \right)$. Hence, the time average of the figure of merit is upper bounded as
\begin{equation}
\label{eq:B00000000008}
{\langle{g(t)}\rangle_{\infty}} \leq \frac{{\|{\omega_{\text{S}}} \|_{\infty}^2}}{{d^{\text{eff}}}({\omega})} ~.
\end{equation}


\begin{figure*}[ht]
\includegraphics[width=1\linewidth]{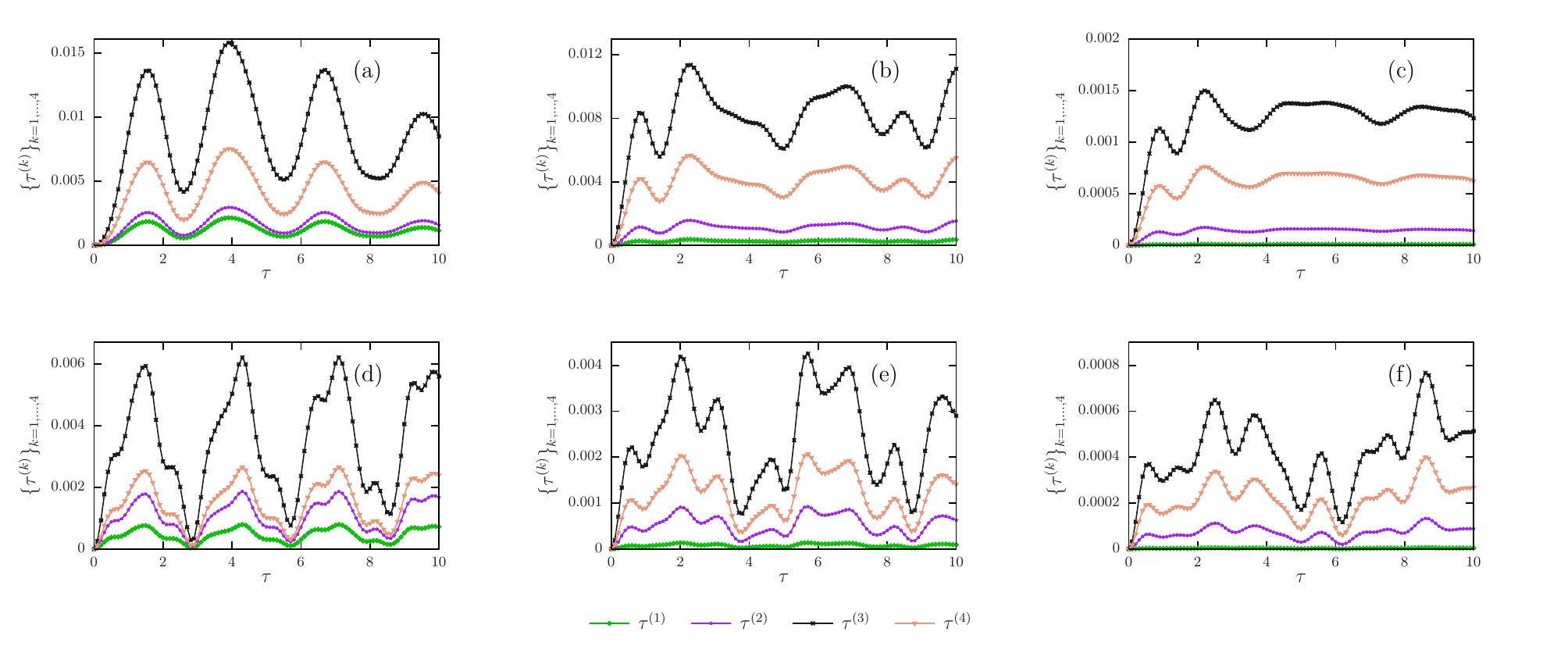}
\caption{(Color online) Plot of the set of QSL times $\{ {\tau^{(k)}} \}_{k = 1,\ldots,4}$ in Eqs.~\eqref{eq:00000000012},~\eqref{eq:00000000016},~\eqref{eq:00000000021} and~\eqref{eq:00000000030}. The panels \ref{figure000004}(a),~\ref{figure000004}(b),~\ref{figure000004}(c) refer to QSL times for the transverse field Ising model [see Eq.~\eqref{eq:00000000032}, with $J = 1$, $h_x = 0.5$, $h_z = -1.05$], and panels \ref{figure000004}(d),~\ref{figure000004}(e),~\ref{figure000004}(f) describe the QSL times for the nonintegrable XXZ model with next-nearest-neighbor hopping [see Eq.~\eqref{eq:00000000033}, with $J = 1$, $U = 2$, ${J_{nnn}} = 0.2$]. The system is initialized at the N\'{e}{e}l state $|\Psi(0)\rangle = |1,0,1,0,\ldots,0,1\rangle$, with open boundary conditions, for the system sizes $L = 4$ [Figs.~\ref{figure000004}(a),~\ref{figure000004}(d)], $L = 6$ [Figs.~\ref{figure000004}(b),~\ref{figure000004}(e)], and $L = 8$ [Figs.~\ref{figure000004}(c),~\ref{figure000004}(f)].}
\label{figure000004}
\end{figure*}

\section{Fluctuations of the purity of the reduced state}
\label{sec:C1}

In this Appendix we follow Ref.~\cite{2010_arxiv_1003.5058} and investigate the role of quantum purity as a figure of merit for equilibration of subsystem $\text{S}$, for which we evaluate the purity ${p_{\text{S}}}(t) = {\text{Tr}_{\text{S}}}({{\rho_{\text{S}}}(t)^2})$, where ${\rho_{\text{S}}}(t) = {\text{Tr}_{\text{B}}}(\rho(t))$ is the reduced density matrix, with $\rho(t) = |\Psi(t)\rangle\langle\Psi(t)|$. The dynamics of the subsystem $\text{S}$ is governed by the equation $d{\rho_{\text{S}}}(t)/dt = i \, {\text{Tr}_{\text{B}}}([\rho(t),H])$, where $H = {H_{\text{S}}}\otimes{\mathbb{I}_{\text{B}}} + {\mathbb{I}_{\text{S}}}\otimes{H_{\text{B}}} + {H_{\text{SB}}}$ is the Hamiltonian of the system. In particular, by exploiting the cyclic property of the trace, $\text{Tr}({A_1}[{A_2},{A_3}]) = \text{Tr}([{A_1},{A_2}]{A_3})$, also using the identities ${\text{Tr}_{\text{SB}}} ( ({\rho_{\text{S}}}(t)\otimes\mathbb{I}_{\text{B}}) [\rho(t),{H_{\text{S}}}\otimes{\mathbb{I}_{\text{B}}}] ) = 0$ and ${\text{Tr}_{\text{SB}}} ( ({\rho_{\text{S}}}(t)\otimes\mathbb{I}_{\text{B}}) [\rho(t),{\mathbb{I}_{\text{S}}}\otimes{H_{\text{B}}}] ) = 0$, one may prove the time derivative of the purity can be written as
\begin{equation}
\label{eq:C00000000001}
\frac{d}{dt}{p_{\text{S}}}(t) = 2 i \, {\text{Tr}_{\text{SB}}} ( ({\rho_{\text{S}}}(t)\otimes\mathbb{I}_{\text{B}}) [\rho(t),{H_{\text{SB}}}] ) ~.
\end{equation}
Next, let ${\rho^{\text{cor}}}(t) = \rho(t) - {\rho_{\text{S}}}(t)\otimes{\rho_{\text{B}}}(t)$ be the traceless correlation operator, i.e., ${\text{Tr}_{\text{S}}}({\rho^{\text{cor}}}(t)) = {\text{Tr}_{\text{B}}}({\rho^{\text{cor}}}(t)) = 0$. This operator is identically zero when the global state $\rho(t)$ is fully uncorrelated, also satisfying the identity ${\text{Tr}_{\text{SB}}} ( ({\rho_{\text{S}}}(t)\otimes\mathbb{I}_{\text{B}}) [{\rho_{\text{S}}}(t)\otimes{\rho_{\text{B}}}(t),{H_{\text{SB}}}] ) = 0$, which implies that
\begin{align}
\label{eq:C00000000002}
\frac{d}{dt}{p_{\text{S}}}(t) = 2 i \, {\text{Tr}_{\text{SB}}} ( ({\rho_{\text{S}}}(t)\otimes\mathbb{I}_{\text{B}}) [{\rho^{\text{cor}}}(t),{H_{\text{SB}}}] ) ~.
\end{align}
In the following, we will derive an upper bound to the rate ${d}{p_{\text{S}}}(t)/dt$ exploiting the role of quantum correlations of the bipartite system $\text{S}+\text{B}$. To do so, note the absolute value of both sides of Eq.~\eqref{eq:C00000000002} can be recast as
\begin{align}
\label{eq:C00000000003}
\left|\frac{d}{dt}{p_{\text{S}}}(t)\right| &\leq 2 \, {\| ({\rho_{\text{S}}}(t)\otimes\mathbb{I}_{\text{B}}) [{\rho^{\text{cor}}}(t),{H_{\text{SB}}}]  \|_1} \nonumber\\
&\leq 2 \, {\| {\rho_{\text{S}}}(t)\otimes\mathbb{I}_{\text{B}} \|_{\infty}} \, {\| [{\rho^{\text{cor}}}(t),{H_{\text{SB}}}]  \|_1}   ~,
\end{align}
where we have used the inequalities $|\text{Tr}({A_1}{A_2})| \leq {\|{A_1}{A_2}\|_1} \leq {\|{A_1}\|_{\infty}}{\|{A_2}\|_1}$, with ${\| {\rho_{\text{S}}}(t)\otimes\mathbb{I}_{\text{B}} \|_{\infty}} = {\| {\rho_{\text{S}}}(t)\|_{\infty}}$. Moreover, invoking the triangle inequality ${\|{A_1} + {A_2}\|_1} \leq {\|{A_1}\|_1} + {\|{A_2}\|_1}$ and also applying the upper bound ${\|{A_1}{A_2}\|_1} \leq {\|{A_1}\|_{\infty}}{\|{A_2}\|_1}$, it follows that $ {\| [{\rho^{\text{cor}}}(t),{H_{\text{SB}}}]  \|_1} \leq 2 \, {\| {\rho^{\text{cor}}}(t) {H_{\text{SB}}} \|_1} \leq 2\, {\| {\rho^{\text{cor}}}(t) \|_1} {\|{H_{\text{SB}}} \|_{\infty}}$. Inserting such results into Eq.~\eqref{eq:C00000000003}, one gets
\begin{equation}
\label{eq:C00000000004}
\left|\frac{d}{dt}{p_{\text{S}}}(t)\right| \leq 4 \, {\| {\rho_{\text{S}}}(t)\|_{\infty}} \, {\| {\rho^{\text{cor}}}(t) \|_1} {\|{H_{\text{SB}}} \|_{\infty}}  ~.
\end{equation}
Interestingly, note the Schatten 1-norm of the correlated operator is upper bounded according to Pinsker's inequality as~\cite{doi:10.1063_1.4871575}
\begin{equation}
\label{eq:C00000000005}
{\| {\rho^{\text{cor}}}(t) \|_1} \leq \sqrt{2\, {I_{\text{SB}}}(\rho(t))} ~,
\end{equation}
where the mutual information is defined
\begin{equation}
\label{eq:C00000000006}
{I_{\text{SB}}}(\rho(t)) := S({\rho_{\text{S}}}(t)) + S({\rho_{\text{B}}}(t)) - S(\rho(t)) ~,
\end{equation}
with $S(\varrho) = - \text{Tr}(\varrho\ln{\varrho})$ being the von Neumann entropy. Hence, by combining Eqs.~\eqref{eq:C00000000004} and~\eqref{eq:C00000000005}, it yields
\begin{equation}
\label{eq:C00000000007}
\left|\frac{d}{dt}{p_{\text{S}}}(t)\right| \leq {4} \, \sqrt{2 {I_{\text{SB}}}(\rho(t))} \, {\| {\rho_{\text{S}}}(t)\|_{\infty}} \, {\|{H_{\text{SB}}} \|_{\infty}}  ~.
\end{equation}

Importantly, Eq.~\eqref{eq:C00000000007} relates the fluctuations of purity of the subsystem $\text{S}$ with the quantum correlations of the whole bipartite system captured by the mutual information. On the one hand, since ${\| {\rho_{\text{S}}}(t)\|_{\infty}} \leq 1$, Eq.~\eqref{eq:C00000000007} implies the upper bound $\left|d{p_{\text{S}}}(t)/dt\right| \leq {4} \, \sqrt{2 {I_{\text{SB}}}(\rho(t))} \, {\| {\rho_{\text{S}}}(t)\|_{\infty}} \, {\|{H_{\text{SB}}} \|_{\infty}}$, which was already presented in Ref.~\cite{PhysRevA.76.042123}. On the other hand, since ${\| {\rho_{\text{S}}}(t)\|_{\infty}} \leq {\| {\rho_{\text{S}}}(t)\|_{2}} = \sqrt{{\text{Tr}_{\text{S}}}({{\rho_{\text{S}}}(t)^2})} =  \sqrt{{p_{\text{S}}}(t)}$, we thus obtain the tighter upper bound
\begin{equation}
\label{eq:C00000000008}
\left|\frac{d}{dt}{p_{\text{S}}}(t)\right| \leq 4 \, \sqrt{2 {I_{\text{SB}}}(\rho(t))} \, {\sqrt{{p_{\text{S}}}(t)}} \, {\|{H_{\text{SB}}} \|_{\infty}}  ~.
\end{equation}
We point out that, since the bipartite system is initialized in a pure state, i.e., $\text{Tr}({\rho(0)^2}) = \text{Tr}({\rho(0)}) = 1$, its instantaneous state $\rho(t) = {U(t)}\rho(0){U^{\dagger}}(t)$ will be pure for all time $t$, i.e., $\text{Tr}({\rho(t)^2}) = \text{Tr}({\rho(t)}) = 1$. As a consequence, the mutual information of the input state tri\-vially collapses into the sum of the von Neumann entropy of the marginal states, ${I_{\text{SB}}}(\rho(t)) = S({\rho_{\text{S}}}(t)) + S({\rho_{\text{B}}}(t)) = 2\, S({\rho_{\text{S}}}(t)) = 2\, S({\rho_{\text{B}}}(t))$. In this case, Eq.~\eqref{eq:C00000000008} can be recast as
\begin{equation}
\label{eq:C00000000009}
\left|\frac{d}{dt}{p_{\text{S}}}(t)\right| \leq {8} \, \sqrt{S({\rho_{\text{S}}}(t))} \, {\sqrt{{p_{\text{S}}}(t)}} \, {\|{H_{\text{SB}}} \|_{\infty}}  ~,
\end{equation}
with the von Neumann entropy satisfying the inequality $0 \leq S({\rho_{\text{S}}}(t)) \leq \ln({d_{\text{S}}})$. Hence, by integrating Eq.~\eqref{eq:C00000000009} over the range $0 \leq t \leq \tau$, it follows that
\begin{equation}
\label{eq:C00000000010}
 \tau \geq \frac{\left| {\sqrt{{p_{\text{S}}}(\tau)} - \sqrt{{p_{\text{S}}}(0)} } \, \right|}{ {4} \, \sqrt{\ln({d_{\text{S}}})}  \, {\|{H_{\text{SB}}} \|_{\infty}} } ~.
\end{equation}
where we have used that $\int dx |g(x)| \geq |\int dx g(x)|$. Here $\tau$ stands for the time to reach the equilibrium purity, i.e., ${p_\text{S}}(\tau) \equiv {p_\text{S}^{\text{eq}}}$. Particularly, if the initial state is fully uncorrelated, i.e., $\rho(0) = {\rho_{\text{S}}}(0)\otimes{\rho_{\text{B}}}(0)$, while ${\rho_{\text{S}}}(0)$ is a pure state with ${p_{\text{S}}}(0) = \text{Tr}({\rho_{\text{S}}}(0)) = 1$, we finally arrive at the bound
\begin{equation}
\label{eq:C00000000011}
 \tau \geq \frac{\left| {\sqrt{{p^{\text{eq}}_{\text{S}}}} - 1} \right|}{ {4} \, \sqrt{\ln({d_{\text{S}}})}  \, {\|{H_{\text{SB}}} \|_{\infty}} } ~.
\end{equation}
We point out Eq.~\eqref{eq:C00000000011} is slightly different from the result presented in Ref.~\cite{2010_arxiv_1003.5058}, where here one finds the equilibration time depends on the square root of the equilibrium purity.


\section{Details on the set of QSL times}
\label{sec:D1}

In this Appendix we provide details on the lower bounds $\{ {\tau^{(k)}} \}_{k = 1,\ldots,4}$ for both the spin models discussed in Sec.~\ref{sec:section000004}. In Figure~\ref{figure000004}, we show the set of QSL times in Eqs.~\eqref{eq:00000000012},~\eqref{eq:00000000016},~\eqref{eq:00000000021} and~\eqref{eq:00000000030} as a function of time $\tau$, for the transverse field Ising model [see Figs.~\ref{figure000004}(a),~\ref{figure000004}(b), and~\ref{figure000004}(c)], and the nonintegrable XXZ model with next-nearest-neighbor hopping [see Figs.~\ref{figure000004}(d),~\ref{figure000004}(e), and~\ref{figure000004}(f)]. In both cases, the spin system is initialized at the N\'{e}{e}l state $|\Psi(0)\rangle = |1,0,1,0,\ldots,0,1\rangle$, with open boundary conditions. Here we set the system sizes $L = 4$ [Figs.~\ref{figure000004}(a),~\ref{figure000004}(d)], $L = 6$ [Figs.~\ref{figure000004}(b),~\ref{figure000004}(e)], and $L = 8$ [Figs.~\ref{figure000004}(c),~\ref{figure000004}(f)]. We see that, for all $\tau \geq 0$ and system size $L$, the curve of $\tau^{(3)}$ is always above the QSL times $\tau^{(1)}$, $\tau^{(2)}$, and $\tau^{(4)}$, regardless of the spin system. This clearly illustrate the fact that ${\tau_{\text{QSL}}} = \max\{\tau^{(1)} , \tau^{(2)} , \tau^{(3)} , \tau^{(4)}\} = \tau^{(3)}$, i.e., the QSL time in Eq.~\eqref{eq:00000000031002aaaa} is given by the lower bound $\tau^{(3)}$, the latter being inversely proportional to the variance of the Hamiltonian governing the dynamics [see Eq.~\eqref{eq:00000000021}]. The quantity $\tau^{(1)}$ stands as a lower bound to the set of QSL times $\{ {\tau^{(2)}}, {\tau^{(3)}}, {\tau^{(4)}} \}$, and approaches small values as the size $L$ grows. In addition, we see the amplitude of $\{ {\tau^{(k)}} \}_{k = 1,\ldots,4}$ decreases as we increase the system size $L$.

\end{document}